\documentclass[12pt]{article}
\usepackage{graphicx,epsfig,psfig,color}
\usepackage{axodraw}
\renewcommand{\theequation}{\arabic{section}.\arabic{equation}}
\newcommand{\bc}{\begin{center}}
\newcommand{\ec}{\end{center}}
\newcommand{\be}{\begin{equation}}
\newcommand{\ee}{\end{equation}}
\newcommand{\bea}{\begin{eqnarray}}
\newcommand{\eea}{\end{eqnarray}}
\newcommand{\ba}{\begin{array}}
\newcommand{\ea}{\end{array}}
\newcommand{\lb}{\label}
\newcommand{\rf}{\ref}
\newcommand{\bfg}{\begin{figure}[htbp]}
\newcommand{\efg}{\end{figure}}
\newcommand{\pr}{Phys.\ Rev.\ }
\newcommand{\np}{Nucl.\ Phys.\ }
\newcommand{\prl}{Phys.\ Rev.\ Lett.\ }
\newcommand{\prp}{Phys.\ Rep.\ }
\newcommand{\ap}{Ann.\ Phys.\ (N.Y.) }
\newcommand{\pl}{Phys. Lett. }
\newcommand{\jmp}{J. Math.\ Phys.\ }
\newcommand{\jp}{J. Phys.\ }

\newcommand{\ptp}{Prog.\ Theor.\ Phys. }

\newcommand{\epj}{Eur.\ Phys.\ J. }
\newcommand{\jhep}{J. High Energy Phys.\ }

\begin{document}

\begin{flushright}
IPNO-DR-06-05
\end{flushright}
\vspace{0.5 cm}
\bc
{\large \textbf{Energy and decay width of the 
\mbox{\boldmath$\pi$}$\mathbf{K}$ atom}}
\vspace{1. cm}

H. Jallouli\\
\textit{Centre National des Sciences et Technologies Nucl\'eaires,\\
Technopole Sidi Thabet 2020, Tunisie\\
E-mail: hassen.jallouli@cnstn.rnrt.tn}\\
\vspace{0.5 cm}
H. Sazdjian\\
\textit{Institut de Physique Nucl\'eaire\footnote
{Unit\'e Mixte de Recherche 8608 du CNRS, de l'IN2P3 et de 
l'Universit\'e Paris XI.},
Groupe de Physique Th\'eorique,\\
Universit\'e Paris XI, F-91406 Orsay Cedex, France\\
E-mail: sazdjian@ipno.in2p3.fr}
\ec
\par
\vspace{1 cm}

\bc
{\large Abstract}
\ec
\par
The energy and the decay width of the $\pi K$ atom are evaluated
in the framework of the quasi\-potential-constraint theory
approach. The main electromagnetic and isospin symmetry
breaking corrections to the lowest-order formulas for the energy 
shift from the Coulomb binding energy and for the decay width are 
calculated. They are estimated to be of the order of a few per cent. 
We display formulas to extract the strong interaction $S$-wave 
$\pi K$ scattering lengths from future experimental data concerning 
the $\pi K$ atom.
\par
\vspace{0.5 cm}
lengths, chiral perturbation theory,
electromagnetic corrections, isospin symmetry breaking.
\newpage

   
\section{Introduction} \lb{s1}
\setcounter{equation}{0}

Hadronic atoms represent new tools for probing strong interaction
dynamics at low energies \cite{dgbth,up,t,bnnt,nm,acs}. 
When the hadronic constituents are 
pseudoscalar mesons, then the probe concerns the properties of
spontaneous chiral symmetry breaking in QCD. The simplest 
representative in that case is the $\pi\pi$ atom (pionium), which
was produced and studied in the DIRAC experiment at CERN 
\cite{dir} and to which many theoretical studies were devoted
\cite{eil,bpt,mrwgo,sil,kur,jsik,lrli,js1,lb,kr,hol,agglr,es}. 
At the next level in
that category one finds the $\pi K$ atom, the properties of which
are tightly related to those of the $SU(3)\times SU(3)$ chiral
symmetry breaking. In particular, the energies and widths of
the atomic levels depend on the strong interaction $\pi K$ 
scattering lengths and through them on various order parameters
of chiral symmetry breaking. The theoretical interest of those
quantities justifies the preparation of new experimental projects
for the production and study of $\pi K$ atoms \cite{dir2}. 
On theoretical grounds, the properties of the $\pi K$ atom were 
recently studied in detail by Schweizer \cite{schw} using the 
approach of nonrelativistic effective field theories to the bound 
state problem \cite{cl1}.
\par
The present work is devoted to the study of the properties of the
$\pi K$ atom using the quasipotential--constraint theory approach
\cite{qp,f,tod,l,js2} which was also applied to the 
pionium case \cite{js1} and which consists of a relativistic and 
covariant three-dimensional formulation of the bound state problem. 
Our aim is to calculate the $O(\alpha)$ corrections to the lowest-order 
formulas for the energy shift and decay width of the $\pi K$ atom, 
where $\alpha$ is the fine structure constant, and to provide the 
means of extracting with sufficient precision the values of the 
strong interaction $S$-wave $\pi K$ scattering lengths from future 
experimental data.
\par
We introduce with respect to our previous
approach to the pionium problem a slight modification in that 
we formulate from the start the bound state problem almost on the
mass shell. This is suggested by the substantial simplification one
gains in order to reach the final results. The leading nonrelativistic 
formulas, which are of order $\alpha^3$, are expressed mainly in terms 
of threshold properties of the strong interaction on-mass shell 
scattering amplitudes. In general,
bound state problems in quantum field theories are formulated in 
terms of off-mass shell scattering amplitudes or kernels; on the 
other hand, effective lagrangians, like the one used in chiral
perturbation theory, give rise to a proliferation of terms, some
of which do not contribute on the mass shell. Therefore, one expects 
many cancellations among unphysical quantities, which are automatically
realized in a formalism based from the start on the use of on-mass 
shell scattering amplitudes and their minimal analytic continuations.
\par
However, because of the presence of infrared divergences, the on-mass
shell formalism ceases to be consistent at higher orders in QED. 
Generally, up to $O(\alpha^4)$ effects in the bound state problem
infrared divergences are unambiguously recognized, isolated and
subtracted or cancelled, according to the method of approach. This 
is an implicit consequence of the Ward identities (and their analogues
for mass terms) satisfied by the theory. At higher orders, for instance
in the treatment of the Lamb shift problem, infrared divergences
can no longer be isolated without ambiguity in an on-mass shell formalism
of the bound state problem. The recourse to an off-mass shell
formalism becomes compulsory at this stage \cite{bbs,fy}. For the present 
problem, since the precision that is sought does not necessitate the 
evaluation of $O(\alpha^5\ln\alpha^{-1})$ effects, this difficulty will 
not show up. 
\par
The main corrections, of order $\alpha$, to the lowest-order formulas
can be represented by three groups of terms: i) the pure 
elecro\-magne\-tic corrections, arising beyond the Coulomb potential;
ii) electromagnetic radiative corrections to the strong interaction
scattering amplitudes, including isospin symmetry breaking effects;
iii) second-order perturbation theory effects in the perturbative
expansion of the bound state energy.
The sum of those corrections is found to be of the order of a few per 
cent relative to the leading terms. Our results agree with those 
obtained by Schweizer \cite{schw}.
\par
The plan of the paper is the following. In Sec. \rf{s2}, we present the
general bound state formalism that we use. In Sec. \rf{s3}, we consider
the specific case of the $\pi K$ atom, for which we adopt a coupled
channel formalism. In Sec. \rf{s4}, we calculate the lowest-order
contributions to the real and imaginary parts of the energy shift.
In Sec. \rf{s5}, pure QED type corrections are evaluated. 
In Sec. \rf{s6}, electromagnetic radiative corrections to the strong
interaction scattering amplitudes are taken into account.  
In Sec. \rf{s7}, a general property of cancellation of divergences of
three-dimensional integrals, present in our formalism, is displayed.
Sec. \rf{s8} deals with second-order effects of the perturbation theory 
expansion of the bound state energy. A summary of results follows in 
Sec \rf{s9}. Some evaluations of integrals are presented in the 
Appendix.
\par 

\section{Bound state formalism} \lb{s2}
\setcounter{equation}{0}

It is generally recognized, on the basis of hamiltonian formalism, 
that relative times and relative energies of particles of
multiparticle systems should not play a dynamical role in relativistic 
theories. Constraint theory \cite{ct1,ct2} allows, through the use of 
first-class constraints, the elimination of redundant variables, 
respecting at the same time the symmetries of the theory (in the present 
case the Poincar\'e invariance). For a two-particle system, described 
by momenta $p_1$ and $p_2$, with physical masses $m_1$ and $m_2$, the 
following constraint
\be \lb{2e1}
C(P,p) \equiv  (p_1^2-p_2^2) - (m_1^2-m_2^2)=0,
\ \ \ \ \ \ P=p_1+p_2,\ \ \ \ p=\frac{1}{2}(p_1-p_2),
\ee
eliminates the relative energy in a covariant way. That constraint
also respects the symmetry between the two particles and remains
valid on the mass shell or in the free case.
\par
We next consider a prototype quantum field theory with a given
lagrangian describing, among others, the interaction between two 
spin-0 particles, 1 and 2. Let $T$ be the on-mass shell elastic scattering
amplitude of the process $1+2\rightarrow 1'+2'$. [$T$ is defined
from the $S$ matrix with the relation $S_{fi}=\delta_{fi}+
(2\pi)^4\delta^4(P-P')T_{fi}$, where $P$ and $P'$ are the total
momenta of the ingoing and outgoing particles.] $T$ is then a function
of the two Mandelstam variables $s=P^2$ and $t=q^2$, with $q=(p_1-p_1')$.
For convenience, we define a modified scattering amplitude $\widetilde T$:
\be \lb{2e2}
\widetilde T(s,t) = \frac{i}{2\sqrt{s}} T(s,t).
\ee
Since the equations we are developing are manifestly covariant, the
total momentum $P$ of the system defines a priviledged direction and
all momenta can be decomposed along longitudinal (one-dimensional)
and transverse (three-dimensional) components with respect to $P$.
However, to simplify notation, we shall henceforth stick to the 
center-of-mass frame ($\mathbf{P=0}$) and represent vectors with their 
temporal and spatial components.
\par
The starting point of the present formalism is the postulate that
$\widetilde T$ satisfies, by means of an effective propagator 
$g_0$, a three-dimensional Lippmann--Schwinger type equation 
leading to the definition of a kernel or a potential $V(s,t)$:
\be \lb{2e3} 
V = \widetilde T - V\,g_0^{}\,\widetilde T.
\ee
This equation can be used to calculate $V$ either iteratively
from $\widetilde T$, or from a perturbative expansion of $\widetilde T$
itself. The iterative integration is three-dimensional taking into
account constraint (\rf{2e1}). Thus, if the two particle momenta
linking $\widetilde T$ to $V$ are $(p_1-k)$ and $(p_2+k)$, total
momentum being conserved, constraint (\rf{2e1}) applied to them
yields $k_0=0$ and integration concerns the three-momentum
$\textbf{k}$. The latter integration does not, however, preserve
the mass-shell character of $\widetilde T$; it forces the corresponding
momentum transfer $-\textbf{k}^2$ to take unphysical values down to
$-\infty$. The situation here is very analogous to that of 
nonrelativistic dynamics. We consider it as corresponding to a minimal 
extension of the scattering amplitude to the off-shell, in the sense that 
it does not imply introduction of new terms, but simply extension of
the domain of validity of the expression of $\widetilde T$ to a larger 
one.
\par
The expression of $g_0^{}$ is chosen so that $V$ is hermitian in the elastic
unitarity region \cite{tod}. Noticing that constraint (\rf{2e1}) implies
equality of the Klein-Gordon operators of particles 1 and 2, 
\bea \lb{2e4}
& &H_0(s,\mathbf{p})\equiv (p_1^2-m_1^2)\big \vert _C  = (p_2^2-m_2^2)\big
\vert _C  = b_0^2(s)-\mathbf{p}^2,\nonumber \\ 
& &b_0^2(s)\equiv \frac {s}{4} - \frac {1}{2} (m_1^2+m_2^2) +
\frac {(m_1^2-m_2^2)^2}{4s},
\eea
$g_0^{}$ is chosen as the propagator associated with these operators:
\be \lb{2e5}
g_0^{}(s,\mathbf{p}) = \frac{1}{H_0(s,\mathbf{p})+i\varepsilon}=
\frac{1}{b_0^2(s)-\mathbf{p}^2+i\varepsilon}.
\ee
On the mass shell, one has $\mathbf{p}^2=b_0^2(s)$ and 
$H_0$ vanishes. However, inside the integration domain of the iterative 
series, $\mathbf{p}$ is replaced by $(\mathbf{p-k})$ and $H_0$ takes the
value $(2\mathbf{p.k}-\mathbf{k}^2)$. 
\par
It is advantageous in some instances to rewrite the
scattering amplitude in terms of a two-particle irreducible kernel.
Introducing the two-particle irreducible kernel $K$ and the
two-particle free propagator $G_0$, one has $T=K+KG_0T$ and
$\widetilde T$ takes the form:
\be \lb{2e5a}
\widetilde T=\widetilde K+\widetilde K\,\widetilde G_0\,\widetilde T,
\ee
where $\widetilde K=iK/(2\sqrt{s})$ and $\widetilde G_0=-i2\sqrt{s}G_0$,
with
\be \lb{2e5b}
G_0=G_{10}\,G_{20},\ \ \ \ \ \ 
G_{a0}=\frac{i}{p_a^2-m_a^2+i\varepsilon},\ \ \ \  a=1,2.
\ee
Then the potential $V$ can be expressed in terms of the 
kernel $\widetilde K$:
\be \lb{2e5c}
V=\widetilde K\Big(1-(\widetilde G_0-g_0)\widetilde K\Big)^{-1},
\ee
the integrations involving $\widetilde G_0$ being four-dimensional.
\par
We assume that $V$ has been calculated (exactly or approximately) from
$\widetilde T$ or $\widetilde K$. Because of the mass shell condition
imposed on the external particles, it is, like $\widetilde T$, a function
of $s$ and $t$ only: $V=V(s,t)$. In $\mathbf{x}$-space, obtained by Fourier
transformation with respect to $\mathbf{q}$ ($=(\mathbf{p}-\mathbf{p}')$),
it is a local function of $\mathbf{x}$; all the nonlocal character of
the interaction is now contained in the energy dependence of $V$. That 
feature does not remain true in an off-mass shell formulation, where
$V$ may depend on the operator $\mathbf{p}^2$.  
\par
We next introduce the Green function associated with $V$ and 
$\widetilde T$ (integrations on internal variables are implicit):
\bea \lb{2e6}
G(s;\mathbf{p},\mathbf{p}')&=&(2\pi)^3\delta^3(\mathbf{p-p'})
g_0^{}(s,\mathbf{p})
+g_0^{}(s,\mathbf{p})\,V(s,-\mathbf{k}^2)\,G(s;\mathbf{p-k},\mathbf{p}')
\nonumber \\
&=&g_0^{}+g_0^{}\,\widetilde T\,g_0^{}.
\eea
Since $G$ is an off-mass shell quantity, the external momenta
$\mathbf{p}$ and $\mathbf{p}'$ are not restricted in the above equations
to the mass shell and $\widetilde T$ is extended off the mass shell
through the continuation of the variable $t$.
\par
In the presence of bound states, $G$ develops poles in $s$. (We assume 
$\mathbf{C}$, $\mathbf{P}$ and $\mathbf{T}$ invariances.) Let $s_0$
be the mass squared of such a bound state ($s_0>0$); then, in the vicinity 
of $s_0$, $G$ behaves as:
\be \lb{2e7}
G_{\stackrel{{\displaystyle \simeq\ }}{s\rightarrow s_0}}
\frac{\Psi\,\Psi^{\dagger}}{(s-s_0+i\varepsilon)},
\ee
where $\Psi^{\dagger}$ is the adjoint of $\Psi$. 
Since (by construction) the kernel $V$ does not have
singularities, at least in the vicinity of the two-body threshold,
one deduces from Eqs. (\rf{2e6}) and (\rf{2e7}) the wave equation
\be \lb{2e8}
\Big[\,g_0^{-1}-V\,\Big]\Psi=0.
\ee
\par
The normalization and orthogonality properties of wave functions are 
obtained in the standard way \cite{f,l,by}. Let us assume that the
spectrum contains several nondegenerate bound states, labelled by an 
integer $n$, with masses squared $s_n>0$. (Generalization to the
degenerate case is straightforward.) Designating by $G^{-1}$ the inverse of 
$G$ [$G^{-1}=g_0^{-1}-V$] and writing the wave equation of the bound state 
$n$ as $G^{-1}(s_n)\Psi_n=0$, one easily arrives at the equation
\be \lb{2e9}
\Psi_n^{\dagger}\left(\frac{G^{-1}(s)-G^{-1}(s_n)}{s-s_n}\right)
G(s)=\frac{\Psi_n^{\dagger}}{s-s_n}.
\ee 
Taking successively the limits $s\rightarrow s_n$ and 
$s\rightarrow s_m$ ($m\neq n$),
one obtains the normalization and orthogonality conditions:
\bea
\lb{2e10}
& &\int\frac{d^3p}{(2\pi)^3}\,
\Psi_n^{\dagger}\Big(\frac{\partial G^{-1}(s)}
{\partial s}\Big)\Big|_{s=s_n}\Psi_n^{}=1,\\ 
\lb{2e11}
& &\int\frac{d^3p}{(2\pi)^3}\Psi_n^{\dagger}
\left(\frac{G^{-1}(s_n)-G^{-1}(s_m)}{s_n-s_m}\right)\Psi_m^{}=0,
\ \ \ \ \ \ s_n\neq s_m.
\eea
\par
The perturbative calculation of energy shifts from a zeroth-order
approximation is also obtained with standard techniques \cite{l,by,agglr}. 
Let us assume that the kernel $V$ can be separated into two parts,
\be \lb{2e12}
V=V_1+V_2,
\ee
so that the solutions corresponding to $V_1$ are known. Let $G_1$
be the Green function associated with $V_1$:
\be \lb{2e13}
G_1=g_0^{}+g_0^{}\,V_1\,G_1.
\ee
We designate by $\varphi_n^{}$ the corresponding bound state wave
functions with masses squared $s_n^{(0)}$. The complete Green function is
constructed from $G_1$ and the kernel $V_2$:
\be \lb{2e14}
G=G_1+G_1V_2G.
\ee
The scattering amplitude due to $V_2$ is obtained from $V_2$ and $G_1$:
\be \lb{2e15}
\widetilde T_2=V_2+V_2\,G_1\,\widetilde T_2,
\ee
from which one deduces:
\be \lb{2e16}
G=G_1+G_1\,\widetilde T_2\,G_1.
\ee
To obtain the perturbative expansion around the bound state $n$,
one isolates the corresponding pole in $G_1$:
\be \lb{2e17}
G_1=G_1'+\frac{\varphi_n^{}\varphi_n^{\dagger}}
{(s-s_n^{(0)}+i\varepsilon)}.
\ee
The reduced scattering amplitude $\widetilde T_2'$ is constructed from 
the reduced Green function $G_1'$ and $V_2$:
\be \lb{2e18}  
\widetilde T_2'=V_2+V_2\,G_1'\,\widetilde T_2'.
\ee
The relationship between $\widetilde T_2$ and $\widetilde T_2'$ is:
\be \lb{2e19}
\widetilde T_2=\left(1-\widetilde T_2'\,
\frac{\varphi_n^{}\varphi_n^{\dagger}}
{(s-s_n^{(0)}+i\varepsilon)}\right)^{-1}\widetilde T_2'.
\ee
Expressing in Eq. (\rf{2e16}) $G_1$ and $\widetilde T_2$ in terms of 
$G_1'$, $\widetilde T_2'$ and the pole term, one obtains the 
decomposition (integrations are implicit)
\be \lb{2e20}
G=G_1'+G_1'\,\widetilde T_2'\,G_1'+\Big(1+G_1'\widetilde T_2'\Big)\,
\frac{\varphi_n\varphi_n^{\dagger}}
{(s-s_n^{(0)}-(\varphi_n^{\dagger}\,\widetilde T_2'\,\varphi_n^{})
+i\varepsilon)}\Big(1+\widetilde T_2'G_1'\Big),
\ee
from which one deduces the value of the mass squared of the bound state $n$
\cite{agglr}:
\be \lb{2e21}
s_n=s_n^{(0)}+(\varphi_n^{\dagger}\,\widetilde T_2'\,\varphi_n^{}).
\ee
Expansion of $\widetilde T_2'$ in terms of $V_2$ and $G_1'$ according to 
Eq. (\rf{2e18}) yields the perturbative series of $s_n$. Up to second 
order in $V_2$, the expression of $s_n$ is \cite{l,by}:
\be \lb{2e22}
s_n=s_n^{(0)}+\bigg\{\,(\varphi_n^{\dagger}\,V_2\,
\varphi_n^{})+(\varphi_n^{\dagger}\,V_2\,G_1'\,V_2\,\varphi_n^{})   
+(\varphi_n^{\dagger}\,V_2\,\varphi_n^{})
(\varphi_n^{\dagger}\,\frac{\partial V_2}{\partial s}\,
\varphi_n^{})\,\bigg\}\bigg|_{s=s_n^{(0)}}.
\ee
\par

\section{\mbox{\boldmath $\pi K$} system} \lb{s3}
\setcounter{equation}{0}

The $\pi K$ atom is formed in the charged sector ($\pi^-K^+$
or its charge conjugate) under the effect of the Coulomb 
interaction and decays under the effect of the strong interaction 
predominantly into the neutral sector ($\pi^0K^0$ or its charge
conjugate). Branching ratios of other decays, involving photons, 
do not exceed a fraction of a per cent. It is natural to 
treat the $\pi K$ system by means of a coupled-channel formalism
by generalizing the formalism developed in Sec. \rf{s2} with a
matrix notation. 
\par
We label with the index $c$ the quantities related to the charged
sector ($\pi^-K^+$) and with the index $n$ those related to the
neutral sector ($\pi^0K^0$). Because of the decay process
$\pi^-K^+\rightarrow \pi^0K^0$, the energy of the bound state
becomes complex with a negative imaginary part. The scattering
amplitudes and Green functions involving the above sectors have 
a common pole at the position of the complex energy of the bound
state. We introduce a two-component wave function $\Psi$ as:
\be \lb{3e1}
\Psi = \left( \ba{c}
\Psi_{c}\\ \Psi_{n} \ea \right)
\ee
and define the potential $V$ in matrix form in the corresponding
space:
\be \lb{3e2}
V = \left( \ba{lc}
V_{cc} & V_{cn}\\
V_{nc} & V_{nn}  \ea \right) .
\ee
\par
The iteration effective propagator $g_0^{}$ [Eqs. (\rf{2e3}),
(\rf{2e4}) and (\rf{2e5})] is now composed of two propagators:
\be \lb{3e3}
g_0^{} = \left( \ba{cc}
g_{0c}^{} & 0\\
0 & g_{0n}^{} \ea \right),\ \ \ \ 
g_0^{-1} = \left( \ba{cc}
g_{0c}^{-1} & 0\\
0 & g_{0n}^{-1} \ea \right) .
\ee
$g_{0c}^{}$ and $g_{0n}^{}$ are defined with the physical masses of
the particles, with respective energy factors $b_{0c}^2(s)$ and
$b_{0n}^2(s)$ [Eq. (\rf{2e4})].
\par
The wave equation (\rf{2e8}) takes now the form of two coupled 
equations:
\bea 
\lb{3e4}
\Big(g_{0c}^{-1}-V_{cc}\Big)\Psi_c-V_{cn}\Psi_n&=&0,\\
\lb{3e5}
-V_{nc}\Psi_c+\Big(g_{0n}^{-1}-V_{nn}\Big)\Psi_n&=&0.
\eea 
The wave function $\Psi_n$ represents an outgoing wave created
by the charged state; it can be eliminated in favor of $\Psi_c$,
yielding the wave equation for the latter wave function:
\be \lb{3e6}
g_{0c}^{-1}\Psi_c=V_{cc}\Psi_c+V_{cn}\Big(1-g_{0n}^{}V_{nn}\Big)^{-1}
g_{0n}^{}V_{nc}\Psi_c.
\ee
This is the bound state equation describing the properties of the
$\pi K$ atom.
\par
The potentials $V$ are calculated from the Lippmann--Schwinger type
equation (\rf{2e3}), written now in matrix form in terms of the 
scattering amplitudes of the processes 
$\pi^-K^+\rightarrow \pi^-K^+$, $\pi^-K^+\rightarrow \pi^0K^0$,
$\pi^0K^0\rightarrow \pi^-K^+$, $\pi^0K^0\rightarrow \pi^0K^0$,
which we designate respectively by $\widetilde T_{cc}$,
$\widetilde T_{nc}$, $\widetilde T_{cn}$ and $\widetilde T_{nn}$,
$\widetilde T$ being defined in Eq. (\rf{2e2}). The relationships of
the components of $V$ and $\widetilde T$ are:
\bea \lb{3e7}
& &V_{cc}=\widetilde T_{cc}-V_{cc}\,g_{0c}^{}\,\widetilde T_{cc}
-V_{cn}\,g_{0n}^{}\,\widetilde T_{nc},\nonumber \\  
& &V_{cn}=\widetilde T_{cn}-V_{cc}\,g_{0c}^{}\,\widetilde T_{cn}
-V_{cn}\,g_{0n}^{}\,\widetilde T_{nn},\nonumber \\  
& &V_{nc}=\widetilde T_{nc}-V_{nc}\,g_{0c}^{}\,\widetilde T_{cc}
-V_{nn}\,g_{0n}^{}\,\widetilde T_{nc},\nonumber \\  
& &V_{nn}=\widetilde T_{nn}-V_{nc}\,g_{0c}^{}\,\widetilde T_{cn}
-V_{nn}\,g_{0n}^{}\,\widetilde T_{nn}.
\eea
When electromagnetism and isospin symmetry breaking are switched-off, 
one remains with the strong interaction or hadronic amplitudes 
$\widetilde T_h^{}$ in the isospin symmetry limit; these are related to 
the isospin invariant amplitudes $\widetilde T^I$ in the following way:
\bea \lb{3e8}
& &\widetilde T_{cc,h}=\frac{1}{3}\Big(2\widetilde T^{1/2}+
\widetilde T^{3/2}\Big),\nonumber \\
& &\widetilde T_{cn,h}=\widetilde T_{nc,h}=
-\frac{\sqrt{2}}{3}\Big(\widetilde T^{1/2}-\widetilde T^{3/2}\Big),
\nonumber \\
& &\widetilde T_{nn,h}=\frac{1}{3}\Big(\widetilde T^{1/2}
+2\widetilde T^{3/2}\Big).
\eea
One also defines the isospin even ($+$) and odd ($-$) amplitudes:
\bea \lb{3e9}
& &\widetilde T^+=\frac{1}{3}\Big(\widetilde T^{1/2}
+2\widetilde T^{3/2}\Big),\nonumber \\
& &\widetilde T^-=\frac{1}{3}\Big(\widetilde T^{1/2}
-\widetilde T^{3/2}\Big).
\eea
The decomposition into partial waves is done according to the 
formula (in the c.m. frame)
\be \lb{3e10}
-2\sqrt{s}\,\widetilde T^I(s,t)=16\pi\sum_{\ell=0}^{\infty}(2\ell+1)
t_{\ell}^I(s)P_{\ell}(\cos\theta),
\ee
while the scattering lengths and effective ranges are defined
from the threshold expansion of the real part of $t_{\ell}^I$:
\be \lb{3e11}
\mathcal{R}e\,t_{\ell}^I(s)=\frac{\sqrt{s}}{2}(\mathbf{p}^2)^{\ell}
\Big(a_{\ell}^I+b_{\ell}^I\mathbf{p}^2+O((\mathbf{p}^2)^2\Big).
\ee
\par
The iteration series (\rf{3e7}) defining the potentials involve
with the presence of the effective propagators $g_0^{}$ 
three-dimensional diagrams, which we call \textit{constraint
diagrams} due to the constraint (\rf{2e1}) that is used there
(see Sec. \rf{s2}). They cancel the $s$-channel 
singularities of the scattering amplitudes in the scattering region 
and provide potentials that are real and regular in the total energy 
variable, being thus appropriate for a continuation to the bound state 
region. 
\par
For Feynman diagrams involving QED parts, the calculation of $V$ can be 
done with a simultaneous perturbative expansion of the scattering 
amplitude in the fine structure constant $\alpha$. Since the bound 
state region is close to the two-particle threshold, one can use
for the evaluation of the magnitude of the corresponding terms the
threshold expansion method developed by Beneke and Smirnov \cite{bs},
which consists of making the expansion already at the level of the
integrand, using dimensional regularization, by recognizing the various
types of infrared singularities that might arise. 
The momenta that are relevant for that type of analysis are 
classified as potential, soft, ultrasoft and hard \cite{bs}. 
\par
In order to apply perturbation theory to Eq. (\rf{3e6}) for the 
evaluation of the bound state energy levels, it is natural to choose 
the nonrelativistic Coulomb potential as the zeroth-order potential,
associated with the nonrelativistic kinetic energy. To this end, we
separate from the total energy $P_0$ the mass term by defining (in the
c.m. frame) the binding energy $\mathcal{E}$:
\be \lb{3e12}
\sqrt{s}=P_0=m_{\pi^-}^{}+m_{K^+}^{}+\mathcal{E}.
\ee
Expanding $s$ with respect to $\mathcal{E}$, one has for the c.m.
momentum squared factor $b_{0c}^2(s)$ [Eq. (\rf{2e4})]:
\be \lb{3e13}
b_{0c}^2(s)\simeq 2\mu \mathcal{E}\left(1+\Big(\frac{m_{\pi^-}^2+m_{K^+}^2
-m_{\pi^-}^{}m_{K^+}^{}}{2m_{\pi^-}^{}m_{K^+}^{}
(m_{\pi^-}^{}+m_{K^+}^{})}\Big)
\mathcal{E}\right),\ \ \ \ \ \ \mu=\frac{m_{\pi^-}^{}m_{K^+}^{}}
{(m_{\pi^-}^{}+m_{K^+}^{})}.
\ee 
The Coulomb potential, which we designate by $V_C^{}$, appears in
the nonrelativistic expansion of the potential $V_{cc}$ with the 
coefficient $2\mu$:
\be \lb{3e14}
V_{cc}=2\mu V_C^{}+\overline V_{cc},\ \ \ \ \ \ V_C^{}=-\frac{\alpha}{r}.
\ee
Therefore, we can divide the whole wave equation by $2\mu$ and recover
at zeroth order the nonrelativistic hamiltonian. The quadratic part
in $\mathcal{E}$ in the expression of $b_0^2(s)$ above can then be 
treated as a part of the perturbation. Furthermore, by absorbing in the
definition of wave functions the factor $(\mu/(m_{\pi^-}+m_{K^+}))^{1/2}$ 
(within the present approximation) one recovers from Eq. (\rf{2e10}) the 
usual nonrelativistic normalization for the zeroth-order wave functions 
$\varphi_n$:
\be \lb{3e15a}
\int\frac{d^3p}{(2\pi)^3}\varphi_n^{\dagger}\varphi_n=1.
\ee
Thus, $\varphi$ satisfies the Schr\"odinger equation
\be \lb{3e15b}
E_n^{(0)}\,\varphi_{n\ell}^m=\Big(\frac{\mathbf{p}^2}{2\mu}+V_C^{}\Big)
\,\varphi_{n\ell}^m,\ \ \ \ \ E_n^{(0)}=-\frac{\mu\alpha^2}{2n^2},
\ \ \ \ \ n\geq \ell+1,
\ee
with $n$, $\ell$ and $m$ representing the principal, orbital and 
azimuthal quantum numbers, respectively.
\par
Similarly, isolating, as in Eq. (\rf{3e12}), from the bound state 
energies the mass term, by writing 
$\sqrt{s_n}=m_{\pi^-}^{}+m_{K^+}^{}+\mathcal{E}_n$ and
$\sqrt{s_n^{(0)}}=m_{\pi^-}^{}+m_{K^+}^{}+E_n^{(0)}$, we recover 
from Eq. (\rf{2e22}) the perturbation series of the nonrelativistic 
theory, written here up to second order:
\bea \lb{3e16}
\mathcal{E}_n&=&E_n^{(0)}+(\frac{1}{2\mu})\bigg\{\,(\varphi_n^{\dagger}
\,V_2\,\varphi_n^{})+(\varphi_n^{\dagger}\,V_2\,G_1'\,V_2\,\varphi_n^{})\,
\bigg\}\bigg|_{\mathcal{E}=E_n^{(0)}}\nonumber\\
& &-\Big(\frac{m_{\pi^-}^2+m_{K^+}^2
-m_{\pi^-}^{}m_{K^+}^{}}{2m_{\pi^-}^{}m_{K^+}^{}
(m_{\pi^-}^{}+m_{K^+}^{})}\Big)E_n^{(0)2}.
\eea
The last term comes from the relativistic corrections of 
the left-hand side of the wave equation (\rf{3e6}) (cf. Eq. (\rf{3e13})).
The last term of Eq. (\rf{2e22}) has been discarded, since it does not
contribute to order $\alpha^4$.
\par
The bound state energy $\mathcal{E}_n$ is in general complex. We 
designate by $E_n$ its real part; its imaginary part is equal to
$-\Gamma_n/2$, where $\Gamma_n$ is the decay width of the bound
state into two neutral mesons. We thus have:
\be \lb{3e16a}
\mathcal{E}_n=E_n-\frac{i}{2}\Gamma_n.
\ee
\par
The Green function $G_1$ [Eq. (\rf{2e13})] that is associated with the 
zeroth-order potential $V_1$, is now essentially the Green function
of the Coulomb potential, which we designate by $G_C^{}$, the expression 
of which has been given by Schwinger \cite{schwg}:
\bea \lb{3e17}  
G_C^{}(E,\mathbf{p},\mathbf{p}')&=&2\mu G_1=
\frac{(2\pi)^3\delta^3(\mathbf{p-p'})}{(E-\mathbf{p}^2/(2\mu))}
-\frac{1}{(E-\mathbf{p}^2/(2\mu))}\,\frac{4\pi\alpha}
{(\mathbf{p-p'})^2}\,\frac{1}{(E-\mathbf{p}^{\prime 2}/(2\mu))}\nonumber\\
& &-\frac{1}{(E-\mathbf{p}^2/(2\mu))}\,4\pi\alpha\eta
I(E,\mathbf{p},\mathbf{p}')\,\frac{1}{(E-\mathbf{p}^{\prime 2}/(2\mu))}
\nonumber \\
&\simeq&2\mu\,\bigg\{\,g_{0c}^{}+g_{0c}^{}2\mu V_C^{}g_{0c}^{}-
g_{0c}^{}8\pi\mu\alpha\eta Ig_{0c}^{}\,\bigg\},
\eea
where $I$ and $\eta$ are defined as
\bea \lb{3e18}
& &I(E,\mathbf{p},\mathbf{p}')=\int_0^1d\rho\,\rho^{-\eta}\,
\Big[\,\rho(\mathbf{p-p'})^2+(1-\rho)^2\frac{\eta^2}{\alpha^2}
(E-\mathbf{p}^2/(2\mu))(E-\mathbf{p}^{\prime 2}/(2\mu))\,\Big]^{-1},
\nonumber \\
& &\eta=\frac{\alpha}{2\sqrt{-E/(2\mu)}}.
\eea
The quantity $G_C'$ will then correspond to $G_C^{}$ from which
the pole term around which perturbative expansion is organized is 
removed: $G_C'=2\mu G_1'$ [Eq. (\rf{2e17})].
\par
The potential $V_2$ that appears in the perturbative formula 
(\rf{3e16}) is obtained from the wave equation (\rf{3e6}), by 
subtracting from the kernel of the right-hand side the Coulomb part 
[Eq. (\rf{3e14})]:
\be \lb{3e19}
V_2=\overline V_{cc}+V_{cn}\Big(1-g_{0n}^{}V_{nn}\Big)^{-1}
g_{0n}^{}V_{nc}.
\ee
\par
In order to use this expression in Eq. (\rf{3e16}), we first expand
the factor $(1-g_{0n}^{}V_{nn})^{-1}$ in $g_{0n}^{}$ and evaluate 
the order of magnitude of each term of the expansion. The integration of 
$g_{0n}^{}$ yields the factor $-i\sqrt{b_{0n}^2(s)}/(4\pi)$
[Eq. (\rf{ae1})], where the total energy is fixed at the bound state 
energy and $b_{0n}^2(s)$ is defined in Eq. (\rf{2e4}) with masses of 
the neutral mesons, $m_{\pi^0}$ and $m_{K^0}$; it is equal to the 
square of the c.m. momentum, which we represent by $p^*$, of the system
$\pi^0K^0$ after the decay of the bound state: $b_{0n}^2(s)=p^{*2}$.
The latter is essentially determined by the mass differences between
charged and neutral mesons. Defining $\Delta m$ as
\be \lb{3e20}
2\Delta m\equiv m_{K^+}^{}-m_{K^0}^{}+m_{\pi^-}^{}-m_{\pi^0}^{}
=0.622\ \mathrm{MeV},
\ee
one has approximately
\be \lb{3e21}
\Big(\frac{p^*}{2\mu}\Big)^2\simeq \frac{\Delta m}{\mu}
\simeq 0.0029,
\ee
which is a quantity of the order of $\alpha/2$. Therefore, each 
$g_{0n}^{}$ can be estimated numerically as a quantity of the order of
$\alpha^{1/2}$. Furthermore, the real part of the energy of the 
bound state receives contributions from even powers of $g_{0n}^{}$,
while the imaginary part (the decay width) receives contributions
from odd powers of $g_{0n}^{}$. With these estimates, Eq. (\rf{3e16})
becomes:
\bea \lb{3e22}
& &2\mu\Delta \mathcal{E}_n=\bigg\{\varphi_{n\ell}^{\dagger}\,
\overline V_{cc}\,\varphi_{n\ell}^{}
+\varphi_{n\ell}^{\dagger}\,V_{cn}\,g_{0n}^{}\,V_{nc}\,\varphi_{n\ell}^{}
\nonumber \\
& &\ \ \ \ \ 
+\varphi_{n\ell}^{\dagger}\,\overline V_{cc}\,\frac{G_C'}{2\mu}\,
\overline V_{cc}\,\varphi_{n\ell}^{}
+\varphi_{n\ell}^{\dagger}\,\Big(\overline V_{cc}\,\frac{G_C'}{2\mu}\,
V_{cn}\,g_{0n}^{}\,V_{nc}+V_{cn}\,g_{0n}^{}\,V_{nc}\,\frac{G_C'}{2\mu}\,
\overline V_{cc}\Big)\varphi_{n\ell}^{}\bigg\}
\bigg|_{\mathcal{E}=E_n^{(0)}}\nonumber \\
& &\ \ \ \ \ -\Big(\frac{m_{\pi^-}^2+m_{K^+}^2
-m_{\pi^-}^{}m_{K^+}^{}}{(m_{\pi^-}^{}+m_{K^+}^{})^2}\Big)E_n^{(0)2}.
\eea
Here, $\Delta \mathcal{E}_n$ represents the energy shift with respect to the
nonrelativistic Coulomb bound state energy (\rf{3e15b}). We have retained, 
for the real part of the energy, terms contributing to the orders $\alpha^3$ 
and $\alpha^4$, while for the imaginary part of the energy (the decay width),
we have retained terms contributing to the orders 
$\alpha^3(\Delta m/\mu)^{1/2}$ and $\alpha^4(\Delta m/\mu)^{1/2}$. 
Higher-order terms are numerically negligible.
\par

\section{Energy shift and decay width in lowest order} \lb{s4}
\setcounter{equation}{0}

We evaluate in this section the lowest-order contributions, $O(\alpha^3)$ 
and $O(\alpha^3(\Delta m/\mu)^{1/2})$, to the real and imaginary parts of 
the energy, respectively; these come from the first two terms of the 
decomposition (\rf{3e22}). Pure electromagnetic interactions beyond the
Coulomb potential contribute only at $O(\alpha^4)$ and hence can be
ignored at the present level. A similar conclusion holds also for
the interference terms between strong and electromagnetic interactions.
Therefore, one has to consider in the potentials $\overline V_{cc}$, 
$V_{cn}$ and $V_{nc}$ solely contributions coming from strong interactions.
\par
All qualitative properties that are derived subsequently depend only
on the short-range nature of the strong interaction, or equivalently,
on the absence of massless particles in it; the particular model used for
representing the strong interaction scattering amplitude is not
relevant, although we will particularly refer to the chiral effective 
lagrangian \cite{w1,gl1,gl2} as a prototype theory
which will also be used for the numerical calculations.
\par
It is preferable here to evaluate $V$ from its relationship with
the two-meson irreducible kernel $\widetilde K$, Eq. (\rf{2e5c}),
which must be considered in its matrix form in the two-channel space.
The kernel $\widetilde K$ is made of vertices and eventually of loops
and may have momentum dependences. By analyzing, with the threshold 
expansion method \cite{bs}, its behavior when integrated in Eq. (\rf{2e5c}) 
with $g_0$ near the two-particle threshold (three-dimensional 
integration), one finds that the eventual three-momentum dependences of 
$\widetilde K$ produce after integration momenta in the form $\mathbf{p}^2$ 
or $\mathbf{q}^2$ multiplying momentum independent pieces. Such terms have 
additional $\alpha^2$ contributions with respect to the momentum
independent pieces when considered in the QED bound states and thus
can be neglected. Concerning the four-dimensional integration with
$\widetilde G_0$, one notices that the presence of loop momenta in the
numerator improves in general the infrared behavior of the integral; in
this case it is only the hard momenta, which feel the ultraviolet 
behavior of the integral, that are concerned by their presence. 
Therefore, without loss of generality, the infrared behavior of integrals
involving the kernel $\widetilde K$ can be studied by considering the
latter as a constant and its value fixed at the two-meson threshold. In 
princple, the total energy is fixed at the bound 
state energy, but since $\widetilde K$ is a smooth function of the energy 
and since the binding energy is of order $\alpha^2$, one can continue 
without harm the energy of $\widetilde K$ up to threshold. By convention, 
one chooses for the strong interaction masses in the isospin symmetry limit 
the charged meson masses.
\par
With the above simplification, the integrations in Eq. (\rf{2e5c}) involve
only $\widetilde G_0$ and $g_0$.  Here, however, the effective propagators
$g_{0c}^{}$ and $g_{0n}^{}$ are defined with the meson physical masses
[Eq. (\rf{2e1})]. It is then necessary, to carry out a consistent 
calculation, to also consider the propagators in $\widetilde G_0$ with 
the corresponding physical masses. The details of the integrations are 
presented in the Appendix [Eqs. (\rf{ae1}), (\rf{ae17}) and (\rf{ae18})]. 
A typical one-loop diagram, involving $\widetilde G_0$, and its constraint 
diagram, involving $g_0$, is presented in Fig. \rf{4f1}.
\par
\bfg
\vspace*{0.5 cm}
\bc
\bc
\begin{picture}(255,75)(0,0)
\SetScale{0.75}

\Line(10,85)(40,50)
\Line(10,15)(40,50)
\Line(140,85)(110,50)
\Line(140,15)(110,50)
\CArc(75,35)(38.079,23.199,156.801)
\CArc(75,65)(38.079,203.199,336.801)

\Line(200,85)(230,50)
\Line(200,15)(230,50)
\Line(330,85)(300,50)
\Line(330,15)(300,50)
\CArc(265,35)(38.079,23.199,156.801)
\CArc(265,65)(38.079,203.199,336.801)
\Text(198.75,37.5)[]{$\times$}

\end{picture}
\ec

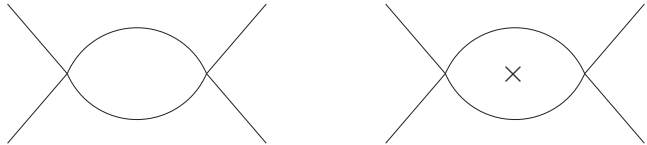
\caption{Two-meson one-loop diagram and its constraint diagram, denoted
with a cross.}
\lb{4f1}
\ec
\efg
The integration of $g_{0c}^{}$, at the bound state energy, provides the
factor $\sqrt{-b_{0c}^2(s)}/(4\pi)$ [Eq. (\rf{ae1})]; it cancels a similar 
factor present in the integral of $\widetilde G_{0c}$ 
[Eqs. (\rf{ae17})-(\rf{ae18})]. Similarly, the inegration of 
$g_{0n}^{}$, at the bound state energy, provides the factor 
$-i\sqrt{b_{0n}^2(s)}/(4\pi)$; it cancels the imaginary part of the 
integral of $\widetilde G_{0n}$. The remaining parts of the 
integrals of $\widetilde G_{0c}$ and $\widetilde G_{0n}$ are real smooth 
functions of the energy and can be replaced, dropping $O(\alpha^2)$ 
corrections, by their values at the charged meson threshold.  
The above calculations can be repeated with all loops present in
Eq. (\rf{2e5c}). One thus recovers the real part of the scattering amplitude
at the charged meson threshold. Once the contributions of the effective 
propagators $g_0^{}$ have been taken into account, and in order to avoid 
double-countings, one must take again in the remaining parts of
the strong interaction amplitudes the equality of the masses of isospin
partners, fixed at the charged meson masses. The effect of mass
differences is taken into account separately within the calculations
of electromagnetic and isospin symmetry breaking contributions.
\par
In summary, the strong interaction potentials that contribute at 
first-order perturbation theory to the complex shift of the bound state
energy are given by the (charged meson) threshold values of the real 
parts of the strong interaction scattering amplitudes:
\bea \lb{4e1} 
\overline V_{cc,h}&=&V_{cc,h}=\mathcal{R}e\,
\widetilde T_{cc,h}\bigg|_{thr.},\nonumber \\
V_{cn,h}&=&V_{nc,h}=\mathcal{R}e\,\widetilde T_{cn,h}
\bigg|_{thr.}=\mathcal{R}e\,\widetilde T_{nc,h}\bigg|_{thr.},
\eea
where the subscript h refers to the purely hadronic part of the 
corresponding quantity. The threshold values of the real parts of the 
scattering amplitudes are given by the $S$-wave scattering lengths 
[Eqs. (\rf{3e8}), (\rf{3e10}) and (\rf{3e11})]. One finds:
\bea \lb{4e2}
V_{cc,h}&=&-\frac{4\pi}{3}(2a_0^{1/2}+a_0^{3/2}),\nonumber \\
V_{cn,h}&=&V_{nc,h}=\frac{4\pi}{3}\sqrt{2}\,(a_0^{1/2}-a_0^{3/2}).
\eea
Since those potentials are constant in momentum space, they yield 
delta-functions in $x$-space and their diagonal matrix elements
become proportional to the square of the modulus of the wave function 
at the origin. In that case, only $S$-wave states contribute. One finds 
for the complex energy shift at first order:
\bea \lb{4e3}
\mathcal{E}_{h,n\ell}^{(1)}&=&
E_{h,n\ell}^{(1)}-\frac{i}{2}\Gamma_{h,n\ell}^{(1)}=
\frac{1}{2\mu}\varphi_{n\ell}^{\dagger}
\Big[\,V_{cc,h}+V_{cn,h}\,g_{0n}^{}\,V_{nc,h}\,\Big]
\varphi_{n\ell}^{}\nonumber \\
&=&-\frac{1}{2\mu}\frac{4\pi}{3}\,\Big[\,(2a_0^{1/2}+a_0^{3/2})+
i\frac{2p_{n0}^*}{3}(a_0^{1/2}-a_0^{3/2})_{}^2\,\Big]\,
\Big|\varphi_{n0}^{}(0)\Big|^2\,\delta_{\ell 0},
\eea
where $p_{n0}^*$ is the c.m. momentum of the neutral mesons
after the decay of the bound state with quantum numbers ($n,\ell=0$).
Taking into account that
\be \lb{4e4}
\Big|\varphi_{n\ell}^{}(0)\Big|^2=\frac{\mu^3\alpha^3}{\pi n^3}
\delta_{\ell 0}
\ee
and introducing the isospin even and odd scattering lengths 
[Eqs. (\rf{3e9})],
\be \lb{4e3a}
a_0^+=\frac{1}{3}(a_0^{1/2}+2a_0^{3/2}),\ \ \ \ \ \ \ 
a_0^-=\frac{1}{3}(a_0^{1/2}-a_0^{3/2}),
\ee
we obtain:
\bea 
\lb{4e5}
E_{h,n0}^{(1)}&=&-2\mu^2\,\frac{\alpha^3}{n^3}\,
\big(a_0^++a_0^-\big),\\
\lb{4e6}
\Gamma_{h,n0}^{(1)}&=&8\,p_{n0}^*\,\mu^2\,\frac{\alpha^3}{n^3}\,
\big(a_0^-\big)_{}^2.
\eea
The above formulas correspond to the expressions found by Deser
\textit{et al.} \cite{dgbth,up,t,bnnt} and provide the leading
effects in the shift in the real part of the energy of the bound
state and in the width of the decay into the neutral mesons.
\par
Numerical predictions about the scattering lengths are made from 
chiral perturbation theory (ChPT). They are summarized in Table 
\rf{4t1}.
\begin{table}[ht]
\bc
\begin{tabular}{|c|c|c|}
\hline
ChPT & $m_{\pi}a_0^{1/2}$ & $m_{\pi}a_0^{3/2}$ \\
\hline
Tree \cite{w2} & 0.14 & $-0.07$ \\
One-loop \cite{bkm} & $0.19\pm 0.02$ & $-0.05\pm 0.02$ \\
Two-loop \cite{bdt} & 0.220 & $-0.047$ \\
\hline
\end{tabular}
\ec
\caption{Theoretical predictions for the $S$-wave scattering
lengths from ChPT.}
\lb{4t1}
\end{table}
\par
On experimental grounds, the values of the scattering lengths
are obtained from an extrapolation of high energy data down to the 
threshold. A complete analysis of the problem, using Roy \cite{r} and 
Steiner \cite{st} equations, has given \cite{bdgm}:
\be \lb{4e7}
m_{\pi}a_0^{1/2}=0.224\pm 0.022,\ \ \ \ \ \ 
m_{\pi}a_0^{3/2}=-(0.448\pm 0.077)\times 10^{-1}.
\ee
\par
Although the two-loop results of ChPT are not yet well understood 
\cite{sk,kkbm}, one notices, with the improvement of the
accuracy of calculations,  a reasonable convergence of the theoretical 
estimates towards the experimental values.
We shall adopt the central values of the experimental results (\rf{4e7})
for the estimates that we shall do in the subsequent sections for the 
various corrections to the lowest-order results; uncertainties in these 
corrections related to the central values will be neglected. 
For the present lowest-order results we obtain for the first three bound 
states the following estimates for the energies, decay widths  and 
lifetimes ($\tau$), presented in Table \rf{4t2}.
\begin{table}[ht]
\bc
\begin{tabular}{|c|c|c|c|c|}
\hline
 $n$ $\ell$ & $E^{(0)}$ (eV) & $E_{h}^{(1)}$ (eV) &
$\Gamma_h^{(1)}$ (eV) & $\tau_{h}^{(1)}$ ($10^{-15}$ s)\\
\hline
 1 0 & $-2898.61$ & $-8.86$ & 0.175 &  3.76 \\
 2 0 &  $-724.65$ & $-1.11$ & 0.022 & 30.09 \\
 2 1 &  $-724.65$ & $\ 0.00$ & 0.000 & $\infty$ \\
\hline
\end{tabular}
\ec
\caption{Zeroth-order energies, first-order energy shifts,
decay widths and lifetimes of hadronic origin of the first 
three bound states.}
\lb{4t2}
\end{table}
\par
In the following, we shall evaluate $O(\alpha)$ corrections to these
results. They originate from three effects: pure electromagnetic 
interaction, electromagnetic radiative corrections to the strong
interaction and second-order perturbation theory effects of the bound
state energy expansion. We shall consider them separately.
\par

\section{Electromagnetic interaction} \lb{s5}
\setcounter{equation}{0}

In this section we consider mainly corrections coming from pure 
electromagnetic interaction. They arise in the channel 
$\pi^-K^+\rightarrow \pi^-K^+$ from one- and two-photon
exchange diagrams and also include vacuum polarization contribution.
The corresponding diagrams, together with the constraint diagram
of the box-ladder diagram, are represented in Fig. \rf{5f1}.
Vertex correction diagrams, associated with self-energy diagrams, 
contribute only at order $\alpha^5\ln\alpha^{-1}$ and may be ignored
at the present level of precision (order $\alpha^4$). 
\par
\bfg
\vspace*{0.5 cm}
\bc
\bc
\begin{picture}(400,180)(0,0)

\Line(10,30)(90,30)
\Line(10,70)(90,70)
\Photon(50,30)(70,70){2}{6.5}
\Photon(50,30)(30,70){-2}{6.5}
\Text(50,10)[]{(e)}

\Line(110,30)(190,30)
\Line(110,70)(190,70)
\Photon(150,70)(170,30){-2}{6.5}
\Photon(150,70)(130,30){2}{6.5}
\Text(150,10)[]{(f)}

\Line(210,30)(290,30)
\Line(210,70)(290,70)
\PhotonArc(270,50)(28.284,135,225){2}{7.5}
\PhotonArc(230,50)(28.284,315,405){2}{7.5}
\Text(250,10)[]{(g)}

\Line(310,30)(390,30)
\Line(310,70)(390,70)
\Photon(350,30)(350,42.5){2}{2.5}
\Photon(350,57.5)(350,70){2}{2.5}
\ArrowArc(350,50)(7.5,90,270)
\ArrowArc(350,50)(7.5,270,450)
\Text(350,10)[]{(h)}

\Line(10,120)(90,120)
\Line(10,160)(90,160)
\Photon(50,120)(50,160){2}{5.5}
\Text(50,100)[]{(a)}

\Line(110,120)(190,120)
\Line(110,160)(190,160)
\Photon(130,120)(130,160){2}{5.5}
\Photon(170,120)(170,160){2}{5.5}
\Text(150,100)[]{(b)}

\Line(210,120)(290,120)
\Line(210,160)(290,160)
\Photon(230,120)(270,160){2}{5.5}
\Photon(270,120)(230,160){2}{5.5}
\Text(250,100)[]{(c)}

\Line(310,120)(390,120)
\Line(310,160)(390,160)
\Photon(330,120)(330,160){2}{5.5}
\Photon(370,120)(370,160){2}{5.5}
\Text(350,140)[]{$\times$}
\Text(350,100)[]{(d)}

\end{picture}
\ec

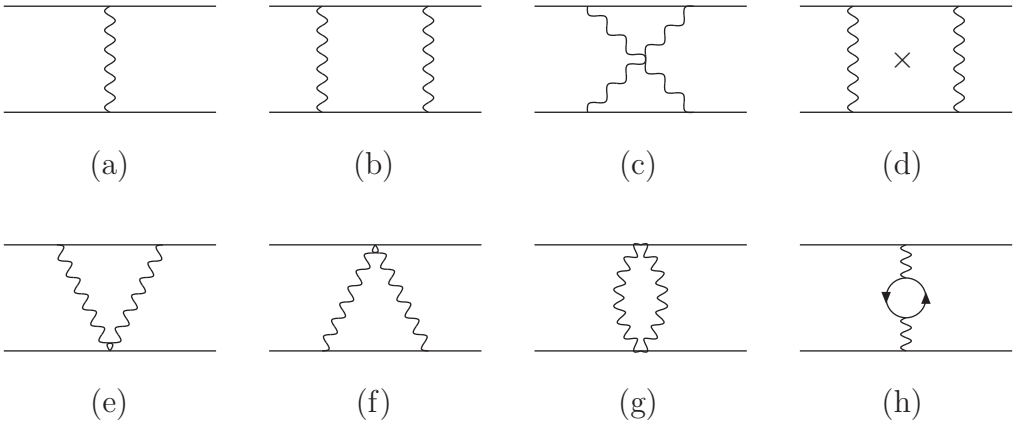
\caption{One- and two-photon exchange diagrams contributing to the
pure electromagnetic corrections. The diagram with a cross represents
the constraint diagram.}
\lb{5f1}
\ec
\efg
The on-mass shell one-photon exchange diagram (a) gives the contribution
\be \lb{5e1}
V_{cc,1\gamma}=\frac{1}{2\sqrt{s}}\frac{e^2}{t}
\Big(2(s-m_{\pi^-}^2-m_{K^+}^2)+t\Big)
\ee
[$e^2=4\pi\alpha$], from which one has to subtract the Coulomb
potential (\rf{3e14}). An expansion of the total energy, according to
Eq. (\rf{3e12}), should also be made.
\par
In the category of two-photon exchange diagrams, we isolate the box-ladder,
crossed-ladder and constraint diagrams, (b), (c), (d). The box-ladder
and crossed-ladder diagrams have separately infrared divergences, as 
well as spurious singularities at threshold. To avoid their appearance,
one must consider the sum of the above three diagrams, within which
several mutual cancellations occur. The mechanism of cancellation is best 
understood with the threshold expansion method \cite{bs}. The leading
part of potential momenta contribution, of order $\alpha^2\ln\alpha^{-1}$,
coming from the box-ladder diagram, is cancelled by that of the constraint 
diagram. The next-to-leading term, of order $\alpha^4$, vanishes in four 
dimensions. Ultra-soft momenta do not contribute on the mass shell. Soft 
momenta contributions of box-ladder and crossed-ladder diagrams, of order
$\alpha^3\ln\alpha^{-1}$, cancel each other. One remains with $O(\alpha^4)$
terms, the sum of which also vanishes on the mass shell. Therefore,
the sum of the three diagrams (b), (c) and (d) does not contribute at
order $\alpha^4$. Details can be found in the Appendix.     
\par
The two-photon exchange diagrams have generally ultraviolet divergences.
The counterterm lagrangian, which is a four-meson contact interaction
term, does not contribute at order $\alpha^4$. 
\par
Diagrams (e) and (f) contribute at order $\alpha^4$. The sum of their
contribution is [Eq. (\rf{ae16})]:
\be \lb{5e2}
V_{cc,2\gamma}=-\frac{(m_{\pi^-}^{}+m_{K^+}^{})}{8\sqrt{s}}
\frac{e^4}{\sqrt{-t}}.
\ee 
\par
Diagram (g) does not contribute at order $\alpha^4$.
\par 
The total corrective contribution of one- and two-photon exchange 
diagrams (without vacuum polarization), $V_{cc,1\gamma}$ minus the 
Coulomb potential (\rf{3e14}) and $V_{cc,2\gamma}$, taking also into 
account the kinematic energy correction factor coming from the 
relativistic wave equation operator (last term of Eq. (\rf{3e22})), is:
\be \lb{5e3}
E_{(1+2)\gamma,n\ell}^{(1)}=\frac{\mu}{8}\Big(3-\frac{\mu}
{(m_{\pi^-}^{}+m_{K^+}^{})}\Big)\frac{\alpha^4}{n^4}
+\frac{\mu^2}{(m_{\pi^-}^{}+m_{K^+}^{})}\frac{\alpha^4}{n^3}
\delta_{\ell 0}-\frac{\mu\alpha^4}{n^3(2\ell+1)}.
\ee
This coincides with similar results obtained from the Bethe--Salpeter
equation (in the Coulomb gauge) \cite{nan} and from the Breit
equation \cite{tod}. 
\par
Another electromagnetic contribution is represented by the vacuum
polarization diagram (h). Generally, in positronium, this diagram
contributes at order $O(\alpha^5)$. However, due to the mass differences
between the electron, entering in the internal loop, and mesons, the 
contribution of the vacuum polarization diagram becomes enhanced \cite{lb}.
It is numerically situated for the energy shift between $O(\alpha^3)$ and
$O(\alpha^4)$ and turns out to be the most important correction to the
strong interaction effect. It can be evaluated analytically \cite{jsik,es}. 
We have evaluated it numerically for the first three bound states, from the 
expression of the corresponding local potential \cite{lp}. The results 
are compatible with the analytic evaluations and are presented below
in Table \rf{5t1}.  
\par 
Apart from the pure electromagnetic corrections, we shall also evaluate 
here the strong interaction contributions to the meson electromagnetic
form factors. The reason is that the interference effects between
strong interaction and electromagnetism in the scattering amplitudes are
evaluated in the literature without the above form factors, which are
considered as parts of the one-photon exchange diagram (see Fig. \rf{5f2}).
\par
\bfg
\vspace*{0.5 cm}
\bc
\bc
\begin{picture}(100,90)(0,0)

\Line(10,20)(90,20)
\Line(10,60)(90,60)
\Photon(50,20)(50,60){2}{5.5}
\GCirc(50,20){5}{0.5}
\GCirc(50,60){5}{0.5}

\end{picture}
\ec

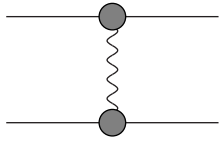
\caption{Strong interaction contribution to the electromagnetic form
factors of mesons.}
\lb{5f2}
\ec
\efg
Defining the hadronic part of the electromagnetic form factor of
pseudoscalar mesons as \cite{gl1} 
\be \lb{5e4}
F_V^{}(t)=1+\frac{1}{6}\langle r^2\rangle t+O(t^2),
\ee
where $\langle r^2\rangle$ defines the mean square charge radius of the meson,
the corresponding contribution through the one-photon exchange 
diagram is:
\be \lb{5e5}
V_{cc,hff}^{}=2\mu e^2\frac{1}{6}\langle r^2\rangle.
\ee
[hff: hadronic form factor.] The sum of the contributions of the 
$\pi$ and $K$ mesons to the energy shift is:
\be \lb{5e6}
E_{hff,n\ell}^{(1)}=\frac{2}{3}\mu^3\,\frac{\alpha^4}{n^3}\,
\Big(\langle r_{}^2\rangle_{\pi^-}^{}+\langle r_{}^2\rangle_{K^+}^{}\Big)
\,\delta_{\ell 0}.
\ee
\par
The mean square radii of the $\pi$ and $K$ mesons were calculated from 
the data in Ref. \cite{bt} in the framework of ChPT to two loops.
The values found there, which we use for the numerical evaluations,
are the following:
\be \lb{5e7}
\langle r_{}^2\rangle_{\pi^-}^{}=(0.452\pm 0.013)\ \mathrm{fm}^2,
\ \ \ \ \ \ \
\langle r_{}^2\rangle_{K^+}^{}=(0.363\pm 0.072)\ \mathrm{fm}^2.
\ee
Numerical estimates of the various electromagnetic contributions
to the first three bound states are summarized in Table \rf{5t1}.
\begin{table}[ht]
\bc
\begin{tabular}{|c|c|c|c|c|}
\hline
 $n$ $\ell$ & $E_{(1+2)\gamma}^{(1)}$ (eV) & $E_{vpol}^{(1)}$ 
(eV) & $E_{hff}^{(1)}$ (eV) & $E_{elm}^{(1)}$ (eV)\\
\hline
 1 0 & $-0.147$ & $-2.561$ & 0.051 & $-2.657$ \\
 2 0 & $-0.025$ & $-0.296$ & 0.006 & $-0.315$ \\
 2 1 & $-0.006$ & $-0.025$ & 0.000 & $-0.031$ \\
\hline
\end{tabular}
\ec
\caption{Electromagnetic corrections (elm) to the energy shift,
composed of contributions of one- and two-photon exchanges 
($(1+2)\gamma$), vacuum polarization (vpol) and hadronic 
form factors (hff).}
\lb{5t1}
\end{table}
\par

\section{Strong interaction in the presence of \protect \\
electromagnetism}
\lb{s6}
\setcounter{equation}{0}

To complete the evaluation of corrections at first order of
perturbation theory, we have now to consider the interference
effects between strong interaction and electromagnetism, including isospin
symmetry breaking. Since we are interested by $O(\alpha)$ corrections
to the strong interaction effects found in Sec. \rf{s4}, it is 
sufficient to consider diagrams with one photon propagator. The analysis 
is best carried out starting from the general relationship of the potential 
$V$ with the two-meson irreducuble kernel $\widetilde K$ [Eq. (\rf{2e5c})], 
which was already used in the pure strong interaction case.
\par
The kernel $\widetilde K$ can be separated into three parts, pure
hadronic, $\widetilde K_h$, pure electromagnetic, $\widetilde K_{\gamma}$, 
and a part with interference between both, $\widetilde K_{h\gamma}$:
$\widetilde K=\widetilde K_h+\widetilde K_{\gamma}+\widetilde K_{h\gamma}$.
Replacing $\widetilde K$ in Eq. (\rf{2e5c}) and then linearizing with 
respect to $\widetilde K_{\gamma}+\widetilde K_{h\gamma}$ and subtracting
the pure hadronic potential and the pure electromagnetic kernel 
$\widetilde K_{\gamma}$, one ends up with the expression of the
interference potential:
\be \lb{6e1}
V_{h\gamma}=\Big(1-\widetilde K_h\,(\widetilde G_0-g_0)\Big)^{-1}\,    
\big(\widetilde K_{\gamma}+\widetilde K_{h\gamma}\big)
\Big(1-(\widetilde G_0-g_0)\,\widetilde K_h\Big)^{-1}-
\widetilde K_{\gamma}.
\ee
Retaining the first few terms, we have for $V_{h\gamma}$ an expansion as
follows:
\be \lb{6e2}
V_{h\gamma}=\widetilde K_{h\gamma}+\widetilde K_h\,(\widetilde G_0-g_0)
\,\widetilde K_{\gamma}+\widetilde K_{\gamma}\,(\widetilde G_0-g_0)\,
\widetilde K_h
+\widetilde K_h\,(\widetilde G_0-g_0)\,
\widetilde K_{\gamma}\,(\widetilde G_0-g_0)\,\widetilde K_h
+\cdots\ .
\ee
In the present approximation, $\widetilde K_{\gamma}$ is the 
one-photon exchange kernel; also $\widetilde K_{h\gamma}$ contains
effects of isospin symmetry breaking; internal propagators, such as
$\widetilde G_0$, should be considered with physical masses.
\par
Typical diagrams, in the charged-charged (cc) channel, are represented 
in Fig. \rf{6f1}.  
\par
\bfg
\vspace*{0.5 cm}
\bc
\bc
\begin{picture}(390,110)(0,0)

\Line(10,30)(80,100)
\Line(10,100)(80,30)
\Photon(27.5,82.5)(62.5,82.5){2}{5.5}
\Text(45,10)[]{(a)}

\Line(110,30)(180,100)
\Line(110,100)(180,30)
\PhotonArc(145,65)(24.749,-45,135){2}{9.5}
\Text(145,10)[]{(b)}

\Line(210,30)(280,100)
\Line(210,100)(280,30)
\Photon(227.5,47.5)(227.5,82.5){2}{5.5}
\Text(245,10)[]{(c)}

\Line(310,30)(380,100)
\Line(310,100)(380,30)
\Photon(327.5,47.5)(327.5,82.5){2}{5.5}
\Text(335,65)[]{$\times$}
\Text(345,10)[]{(d)}

\end{picture}
\ec

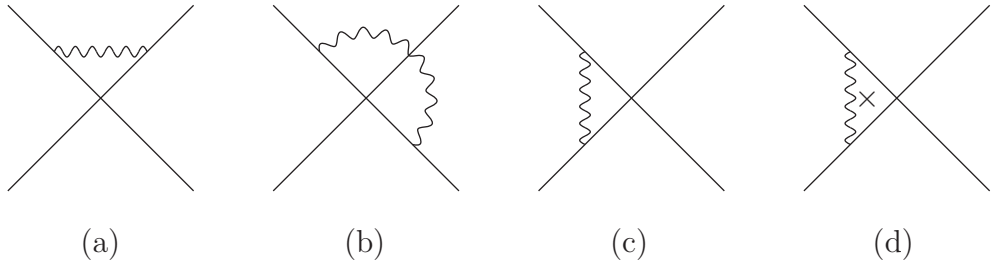
\caption{Electromagnetic radiative corrections to the strong interaction
in the charged-charged channel; diagram (d) is the constraint diagram
of diagram (c); symmetric diagrams to those, as well as self-energy 
diagrams, are not drawn.}
\lb{6f1}
\ec
\efg
Diagrams (a), (b), (c) represent one-photon exchanges in the $t$-channel,
$u$-channel and $s$-channel respectively; diagram (d) is the constraint
diagram of diagram (c). Also four other diagrams symmetric to the above 
ones exist; futhermore, one must also include the contributions of the
self-energy diagrams. Each of those diagrams has infrared divergences on
the mass shell; however, mutual cancellations occur by grouping several
diagrams. The sum of the $t$-channel diagrams and the contributions of
the self-energy diagrams is finite at order $\alpha^4$. The constraint
diagram cancels the infrared divergence and spurious singularities of
the $s$-channel diagram generated by the potential momenta \cite{bs}.
The remaining part of the $s$-channel diagram, which is still infrared
divergent and generated by the hard momenta, is associated with the
$u$-channel diagram, which has similar singularities; their sum is
free of divergences and of spurious singularities. Diagrams where the
photon is emitted from the vertex (not drawn in Fig. \rf{6f1}) are not 
infrared singular and contribute as $O(\alpha^4)$. The sum of all the
above diagrams is infrared finite and of order $\alpha^4$. It defines the
regularized real part of the strong interaction vertex in the presence
of electromagnetism; its value at the bound state energy differs from
its value at threshold by an $O(\alpha^2)$ term and therefore can be
replaced by the value of the regularized real part of the vertex at 
threshold. (See Appendix for details.)
\par
In the charged-neutral (cn) or neutral-charged (nc) channels, $t$-channel 
and $u$-channel type diagrams are
absent. In that case, one has to associate the part of the $s$-channel
diagram generated by the hard momenta with the self-energy contributions
of the external charged particles and the same type of cancellations as
above operate, leading to the same qualitative result.
\par
The analysis done above can be repeated with more complicated diagrams.
One has always to group several diagrams of the same class to reach,
with the aid of the threshold expansion method \cite{bs}, mutual
cancellations of infrared divergences and spurious singularities.
The final results that we obtain are very similar to those found
in Sec. \rf{s4}, Eqs. (\rf{4e1}), with the only difference that the 
pure strong interaction scattering amplitude is now replaced by the
the regularized strong interaction amplitude in the presence of 
electromagnetism and isospin symmetry breaking:
\bea \lb{6e3}
\overline V_{cc,h+h\gamma}^{}&=&V_{cc,h+h\gamma}=\mathcal{R}e\,
\widetilde T_{cc,h+h\gamma}^{reg.}\bigg|_{thr.},
\nonumber \\
V_{cn,h+h\gamma}^{}=V_{nc,h+h\gamma}^{}&=&V_{cn}^{}=V_{nc}^{}=
\mathcal{R}e\,\widetilde T_{cn}^{reg.}
\bigg|_{thr.}=\mathcal{R}e\,\widetilde T_{nc}^{reg.}\bigg|_{thr.}.
\eea
The threshold values of the regularized real parts of the strong
interaction scattering amplitudes in the presence of electromagnetism
deviate from the strong interaction $S$-wave scattering lengths 
[Eqs. (\rf{4e2})] by small amounts that we designate by 
$(\Delta(a_0^++a_0^-))_{h\gamma}$ and $(\Delta a_0^-)_{h\gamma}$. 
One then has:
\bea \lb{6e4}
V_{cc,h+h\gamma}&=&-4\pi\Big((a_0^++a_0^-)+
(\Delta(a_0^++a_0^-))_{h\gamma}\Big),\nonumber \\
V_{cn}^{}&=&4\pi\sqrt{2}\Big(a_0^-+(\Delta a_0^-)_{h\gamma}\Big),
\eea
where $a_0^+$ and $a_0^-$ are the pure strong interaction 
isospin even and odd scattering lengths, respectively [Eqs. (\rf{4e3a})].
For further use, we define the following relative amounts:
\be \lb{6e5a}
\delta_{cc,h\gamma}^{(1)}\equiv \frac{(\Delta(a_0^++a_0^-))_{h\gamma}}
{(a_0^++a_0^-)},\ \ \ \ \ \ 
\delta_{cn,h\gamma}^{(1)}\equiv \frac{(\Delta a_0^-)_{h\gamma}}{a_0^-}.
\ee
\par
The various diagrams entering in the calculation of the strong
interaction scattering amplitude in the presence of electromagnetism 
have also ultraviolet divergences. They are eliminated by the 
low energy constants of the effective chiral lagrangian in the
presence of electromagnetism \cite{ur,ku}. 
\par
Turning back to the infrared problem, we emphasize the following point.
The constraint diagrams have eliminated, to order $\alpha^4$, the 
infrared divergences and threshold singularities of the scattering
amplitudes and allowed the definition of regularized real parts of them.
In the literature, the scattering amplitudes are generally calculated
in the scattering region; then, infrared divergences are eliminated
by factorizing the Coulomb phase \cite{lr},
which does not contribute to the cross section, and by combining the
process under consideration with real soft photon emission processes
\cite{ku,nt,km}. The remaining threshold singularities are then subtracted 
to define a regularized scattering amplitude at threshold. In the present 
formalism, the use of constraint diagrams for the definition of the 
potentials circumvents the latter procedures on the mass shell (to order 
$\alpha^4$), providing directly the regularized result. It can be checked 
explicitly that the pieces that are cancelled by the constraint diagrams
are the same quantities that are subtracted in procedures dealing with
the scattering amplitudes in the scattering region. Therefore, the
regularized real parts of the amplitudes that we have defined in Eqs.
(\rf{6e3}) are identical to those defined in the literature.
\par
More complicated diagrams than those of Fig. \rf{6f1} involve an 
increasing number of loops. However, in ChPT, the increase in the number
of loops decreases the order of magnitude of the corresponding correction
at low energy; this is why the chiral perturbation theory is organized
in terms of the number of loops \cite{w1,gl1,gl2}. Therefore, the diagrams 
of the type of Fig. \rf{6f1} represent the most important contributions to 
the interference effects between strong interaction and electromagnetism,
apart from isospin partner mass difference insertion terms at the tree 
level, and practical calculation of these effects have been limited to them
\cite{ku,nt,km}. Next-to-leading effects are represented by diagrams 
of the type of Fig. \rf{6f2}.   
The latter diagrams contain infrared logarithmic contributions which 
partially enhance their order of magnitude and will be taken into
account in Sec. \rf{s8} in conjunction with second-order perturbation
theory effects.
\par
\bfg
\vspace*{0.5 cm}
\bc
\bc
\begin{picture}(400,100)(0,0)

\Line(10,30)(20,60)
\Line(10,90)(20,60)
\Line(90,30)(80,60)
\Line(90,90)(80,60)
\CArc(50,50)(31.622,18.435,161.565)
\CArc(50,70)(31.622,198.435,341.565)
\Photon(50,38.378)(50,81.622){2}{6.5}
\Text(50,10)[]{(a)}

\Line(110,30)(120,60)
\Line(110,90)(120,60)
\Line(190,30)(180,60)
\Line(190,90)(180,60)
\CArc(150,50)(31.622,18.435,161.565)
\CArc(150,70)(31.622,198.435,341.565)
\Photon(150,38.378)(150,81.622){2}{6.5}
\Text(165,60)[]{$\times$}
\Text(150,10)[]{(b)}

\Line(210,30)(220,60)
\Line(210,90)(220,60)
\Line(290,30)(280,60)
\Line(290,90)(280,60)
\CArc(250,50)(31.622,18.435,161.565)
\CArc(250,70)(31.622,198.435,341.565)
\Photon(250,38.378)(250,81.622){2}{6.5}
\Text(238,60)[]{$\times$}
\Text(250,10)[]{(c)}

\Line(310,30)(320,60)
\Line(310,90)(320,60)
\Line(390,30)(380,60)
\Line(390,90)(380,60)
\CArc(350,50)(31.622,18.435,161.565)
\CArc(350,70)(31.622,198.435,341.565)
\Photon(350,38.378)(350,81.622){2}{6.5}
\Text(338,60)[]{$\times$}
\Text(365,60)[]{$\times$}
\Text(350,10)[]{(d)}

\end{picture}
\ec

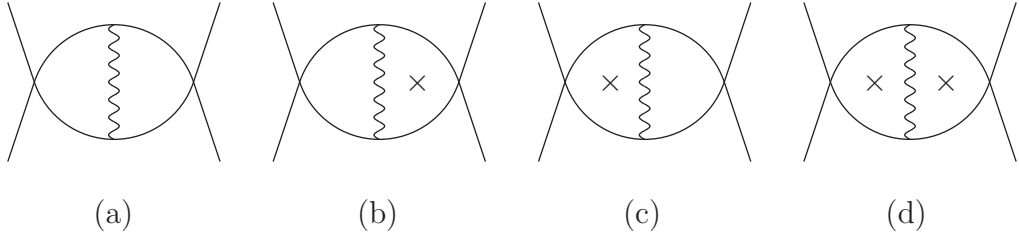
\caption{Two-loop diagram and its constraint diagrams.}
\lb{6f2}
\ec
\efg
The evaluation for the $\pi K$ atom of contributions of diagrams of the
type of Fig. \rf{6f1} and of mass shift insertion effects  have been done 
by two groups of authors \cite{nt,km}. Although their results do not 
exactly coincide, the numerical values that emerge for the corrective 
terms are close to each other. We use here the numerical values given in 
Ref. \cite{km}. One has:
\bea \lb{6e5}
\delta_{cc,h\gamma}^{(1)}&=&O(e^2)+O(e^2\,p^2)
+O\big(\,|m_u-m_d|\,p^2\,\big)\nonumber \\
&=&\Big(1.2-(0.3\pm 3.2)+0.2\Big)\times 10^{-2}
=\big(1.1\pm 3.2\big)\times 10^{-2}, 
\eea
\bea \lb{6e6}
\delta_{cn,h\gamma}^{(1)}&=&O(e^2)+O\big(\,|m_u-m_d|\,\big)+
O(e^2\,p^2)+O\big(\,|m_u-m_d|\,p^2\,\big)\nonumber \\
&=&\Big(0.8+0.5-(0.8\pm 0.7)+(0.7\pm 0.2)\Big)\times 10^{-2}
=\big(1.2\pm 0.7\big)\times 10^{-2},\nonumber \\
& &
\eea     
where $O(e^2)$ and $O\big(\,|m_u-m_d|\,\big)$ represent effects of
electromagnetism and isospin symmetry breaking mass insertions at the 
tree level, $O(e^2\,p^2)$ and $O\big(\,|m_u-m_d|\,p^2\,\big)$ effects 
of electro\-magne\-tism and isospin symmetry breaking at one-loop level; 
the uncertainties arise mainly from the electro\-magne\-tic low energy 
constants of the effective chiral lagrangian, which are estimated by 
their order of magnitude.
\par
The above modifications of threshold values of the amplitudes modify the
results (\rf{4e5}) and (\rf{4e6}), obtained in the pure strong interaction 
case, with the inclusion of the corresponding corrections:
\bea 
\lb{6e7}
E_{h+h\gamma,n0}^{(1)}&=&-2\mu^2\,\frac{\alpha^3}{n^3}\,
\big(a_0^++a_0^-)\big)\,\big(1+\delta_{cc,h\gamma}^{(1)}\big)
=E_{h,n0}^{(1)}\,\big(1+\delta_{cc,h\gamma}^{(1)}\big)\nonumber \\
&=&E_{h,n0}^{(1)}\,\Big(1+(1.1\pm 3.2)\times 10^{-2}\Big),\\
\lb{6e8}
\Gamma_{h+h\gamma,n0}^{(1)}&=&8\,p_{n0}^*\,\mu^2\,\frac{\alpha^3}{n^3}\,
\big(a_0^-\big)_{}^2\,\big(1+2\delta_{cn,h\gamma}^{(1)}\big)
=\Gamma_{h,n0}^{(1)}\,\big(1+2\delta_{cn,h\gamma}^{(1)}\big)
\nonumber \\
&=&\Gamma_{h,n0}^{(1)}\,\Big(1+(2.4\pm 1.4)\times 10^{-2}\Big).
\eea
\par

\section{Cancellation of divergences of three-dimensional diagrams}
\lb{s7}
\setcounter{equation}{0}

We shall show in this section that divergences of constraint diagrams
do not appear in physical quantities. 
Up to order $\alpha^4$ effects, the only divergences that are
introduced by constraint diagrams are those that cancel the 
infrared divergences of the on-mass shell scattering amplitudes.
Other divergences that might appear in the perturbation series of
the energy shifts are actually mutually cancelled. 
\par
To observe the latter property, we go back to the matrix notation and 
express directly the scattering amplitude $\widetilde T_2'$ 
[Eq. (\rf{2e18})], that defines the perturbation expansion of the energy 
shifts [Eq. (\rf{2e21})], in terms of two-particle irreducible kernels.
Let $\widetilde K_1$ be the two-particle irreducible kernel of the
unperturbed theory, that defines potential $V_1$; we have:
\be \lb{7e1}
V_1=\widetilde K_1\Big(1-(\widetilde G_0-g_0)\widetilde K_1\Big)^{-1}.
\ee
\par
Potential $V_2$ is defined as $V_2=V-V_1$, where $V$ is the the 
potential corresponding to the total interaction. It is related to
the total scattering amplitude $\widetilde T$ [Eq. (\rf{2e3})] and 
to its two-particle irreducible kernel $\widetilde K$ [Eq. (\rf{2e5c})], 
which in turn can be separated into two parts, after isolating in it 
$\widetilde K_1$:
\be \lb{7e2}
\widetilde K=\widetilde K_1+\widetilde K_2.
\ee
For the Green function $\widetilde G_1'$ that enters in the definition
of $\widetilde T_2'$ [Eq. (\rf{2e18})], one can use a decomposition 
similar to that of Eq. (\rf{3e17}):
\be \lb{7e3}
G_1'=g_0^{}+g_0^{}V_1g_0^{}+G_1'',
\ee
where $G_1''$ corresponds to the part created by
multiparticle or multiphoton exchanges.
\par 
Replacing then $V_1$ and $V$ in the expression of $\widetilde T_2'$ 
in terms of $\widetilde K_1$ and $\widetilde K$, respectively, and
iterating $\widetilde T_2'$ and $\widetilde G_1'$, one obtains for
$\widetilde T_2'$ the following expansion, in which we have kept up
to two propagator terms:
\bea \lb{7e4}
\widetilde T_2'&=&\widetilde K_2+\widetilde K_2\widetilde G_0\widetilde K_2
+\widetilde K_2\widetilde G_0\widetilde K_2\widetilde G_0\widetilde K_2 
+\widetilde K_2\widetilde G_0\widetilde K_1\widetilde G_0\widetilde K_2   
\nonumber \\
& &+\widetilde K_2\widetilde G_1''\widetilde K_2
+\widetilde K_1(\widetilde G_0-g_0^{})\widetilde K_2
+\widetilde K_2(\widetilde G_0-g_0^{})\widetilde K_1
\nonumber \\
& &+\widetilde K_1(\widetilde G_0-g_0^{})\widetilde K_2
\widetilde G_0\widetilde K_2
+\widetilde K_2\widetilde G_0\widetilde K_2
(\widetilde G_0-g_0^{})\widetilde K_1\nonumber \\
& &+\widetilde K_1(\widetilde G_0-g_0^{})\widetilde K_1
(\widetilde G_0-g_0^{})\widetilde K_2
+\widetilde K_2(\widetilde G_0-g_0^{})\widetilde K_1
(\widetilde G_0-g_0^{})\widetilde K_1\nonumber \\
& &+\widetilde K_1(\widetilde G_0-g_0^{})\widetilde K_2
(\widetilde G_0-g_0^{})\widetilde K_1+\cdots \ .
\eea
\par
The important point to be noticed is that the effective propagator
$g_0^{}$ is completely absent in terms containing only the kernel 
$\widetilde K_2$. $g_0^{}$ is present in terms that contain 
$\widetilde K_1$ on their left or right boundaries. 
Since the higher-order electromagnetic corrections with respect to
the Coulomb potential appear at order $\alpha^4$ only at first order 
of perturbation theory, one may simplify the analysis by assuming
that $\widetilde K_1$ corresponds essentially to the pure
elecro\-mag\-ne\-tic part, $\widetilde K_{\gamma}$, and $\widetilde K_2$ 
to the pure hadronic part, $\widetilde K_h$, as well as to the 
interference part of the hadronic and electro\-mag\-ne\-tic 
interactions, $\widetilde K_{h\gamma}$.
The terms $\widetilde K_1\widetilde G_0\widetilde K_2$ 
or $\widetilde K_2\widetilde G_0\widetilde K_1$, in which $\widetilde K_2$ 
is represented by a constant contact term, have 
infrared divergences, the dominant part of which is cancelled by the
constraint propagator $g_0^{}$; the non-dominant divergences are
mutually cancelled, by considering together self-energy diagrams
and eventually other combinations of $t$-channel and $u$-channel
diagrams (cf. Sec. \rf{s6}). The term 
$\widetilde K_2\widetilde G_0\widetilde K_1\widetilde G_0\widetilde K_2$, 
which in lowest order is represented by diagram (a) of Fig. \rf{6f2}, is
finite and actually its constraint diagrams (b), (c) and (d) of
Fig. \rf{6f2} have been cancelled by similar terms that have 
appeared through the perturbative expansion of the bound state energy. 
Similarly, the
constraint diagrams which were present in the definition of the
hadronic potentials $V_h$ (Sec. \rf{s3}) also disappeared. Those
diagrams, considered individually, contain ultraviolet or infrared
divergences (linear or logarithmic). The present cancellation 
mecanism shows that their effect is irrelevant. However, the 
way of organizing the perturbation expansion in terms of the
potentials $V$, rather than in terms of the kernels $\widetilde K$ or 
the amplitudes $\widetilde T$, has the advantage, through cancellation
effects by constraint diagrams, of naturally arriving at quantities
that are smooth functions of the energy in the vicinity of the
two-particle threshold, being continued, within the present 
approximations, to threshold (cf. Eqs. (\rf{4e1}) and (\rf{6e3})).
Actually, it is only the finite parts of their contributions that are 
relevant for that operation. This is why, we shall continue 
formulating the perturbative expansion of the bound state energy in 
terms of potentials, keeping in mind that in mutually cancelling constraint
diagrams the same convention should be used when removing divergences.
\par
In summary, no ultraviolet or infrared divergences globally occur from 
three-dimen\-sional diagrams in physical quantities up to oreder
$\alpha^4$.
\par

\section{Second order of perturbation theory} \lb{s8}
\setcounter{equation}{0}

We evaluate, in this section, the contributions of second-order 
perturbation theory effects, which are represented by the terms
containing the subtracted Coulomb Green function $G_C'$ in
Eq. (\rf{3e22}). Here, the potentials that have significant 
contributions are the hadronic part  and the vacuum polarization 
part of $\overline V_{cc}$. 
\par
We first consider the hadronic part $V_{cc,h}$ of $\overline V_{cc}$,
which was already studied in Sec. \rf{s4}. Its contribution to the
energy shift at second-order of perturbation theory is represented by 
the following sum of terms:
\be \lb{8e1}
2\mu \mathcal{E}_{hh,n\ell}^{(2)}=\varphi_{n\ell}^{\dagger}\,\Big[\,
V_{cc,h}\,\frac{G_C'}{2\mu}\,V_{cc,h}+V_{cc,h}\,\frac{G_C'}{2\mu}\,
V_{cn,h}\,g_{0n}^{}\,V_{nc,h}+V_{cn,h}\,g_{0n}^{}\,V_{nc,h}\,
\frac{G_C'}{2\mu}\,V_{cc,h}\,\Big]\,\varphi_{n\ell}.
\ee
(A negligible contribution, involving two $g_{0n}^{}$s, has been
omitted.)
Since at the approximation we are working $V_{cc,h}$ and $V_{cn,h}$
are constants in momentum space [Eqs. (\rf{4e2})], the contribution of
$G_C'$ factorizes with its integrations, and the wave functions become
projected on their values at the origin in $x$-space:
\be \lb{8e2}  
2\mu \mathcal{E}_{hh,n\ell}^{(2)}=V_{cc,h}\,\Big[\,\int\,
\frac{d^3p}{(2\pi)^3}\frac{d^3p'}{(2\pi)^3}\frac{1}{2\mu}
G_C'(E_{n\ell}^{(0)},\mathbf{p},\mathbf{p}')\,\Big]\,
|\varphi_{n\ell}(0)|^2\,\Big[\,V_{cc,h}+
2V_{cn,h}\,g_{0n}^{}\,V_{nc,h}\,\Big].
\ee
Defining
\bea 
\lb{8e3}
& &\langle \frac{G_C'}{2\mu}\rangle_{n\ell}^{}\equiv
\int\,\frac{d^3p}{(2\pi)^3}\frac{d^3p'}{(2\pi)^3}\,
\frac{1}{2\mu}G_C'(E_{n\ell}^{(0)},
\mathbf{p},\mathbf{p}'),\\
\lb{8e4}
& &\delta_{hh,n\ell}^{(2)}\equiv V_{cc,h}\,
\langle \frac{G_C'}{2\mu}\rangle_{n\ell}^{},
\eea
and taking into account results (\rf{4e2}), (\rf{4e3}), (\rf{4e5})
and (\rf{4e6}) one obtains:
\be \lb{8e5}
\mathcal{E}_{hh,n\ell}^{(2)}=E_{hh,n\ell}^{(2)}-\frac{i}{2}
\Gamma_{hh,n\ell}^{(2)}=\delta_{hh,n0}^{(2)}\,\Big(E_{h,n0}^{(1)}-
i\Gamma_{h,n0}^{(1)}\Big)\,\delta_{\ell 0}^{},
\ee
which in turn yields:
\be \lb{8e6}
\frac{E_{hh,n0}^{(2)}}{E_{h,n0}^{(1)}}=\delta_{hh,n0}^{(2)},\ \ \ \ \ 
\frac{\Gamma_{hh,n0}^{(2)}}{\Gamma_{h,n0}^{(1)}}=2\delta_{hh,n0}^{(2)}.
\ee
\par
The calculation therefore amounts to that of the double integral of 
$G_C'/(2\mu)$. The latter is composed of three contributions 
[Eq. (\rf{3e17})].
\par
The first corresponds to zero-photon exchange and its integral is
equal to that of $g_{0c}^{}$, i.e., to $\sqrt{-b_{0c}^2(s)}/(4\pi)$
[Eq. (\rf{ae1})].
\par
The second term corresponds to one-photon exchange; its integral is
ultraviolet divergent, but this divergence is cancelled by that of
the three constraint diagrams of Fig. \rf{6f2} (see Appendix for details).
The finite part of the latter in turn cancels a finite logarithmic piece of
the four-dimensional diagram (a) of Fig. \rf{6f2}; therefore, the finite
part of the sum of the four diagrams of Fig. \rf{6f2} becomes a smooth
function. The finite part of the one-photon exchange part of the
integral of $G_C'/(2\mu)$ simply isolates the dominant logarithmic part
of the four-dimensional diagram.
\par
The third term corresponds to the multiphoton exchanges and is finite.
It can be calculated in several ways: either using an integration by parts 
in the variable $\rho$ and isolating first the pole term to be subtracted 
\cite{cl2}, or integrating first with respect to the momenta and isolating 
at the end the pole term. 
\par
The result is, for the finite part, using dimensional regularization and 
writing the contributions of the above three terms in successive order:
\bea \lb{8e7}
\langle \frac{G_C'}{2\mu}\rangle_{n0}^{}
&=&\frac{\mu\alpha}{4\pi}\frac{1}{n}-\frac{\mu\alpha}{2\pi}
\ln(\frac{n}{\alpha})+\frac{\mu\alpha}{2\pi}\,\Big(\psi(n)-\psi(1)
-\frac{3}{2n}\,\Big),
\eea
where $\psi$ is the logarithmic derivative of the Gamma function.
\par  
Numerically, one finds, using Eq. (\rf{4e2}):
\be \lb{8e8}
\delta_{hh,10}^{(2)}=0.009,\ \ \ \ \ \delta_{hh,20}^{(2)}=0.008.
\ee
\par
We next consider the interference term between the hadronic and vacuum 
polarization parts of $\overline V_{cc}$. The corresponding energy shift
is:
\be \lb{8e9}
2\mu \mathcal{E}_{hvpol,n\ell}^{(2)}=2\varphi_{n\ell}^{\dagger}\,
V_{cc,vpol}\,\frac{G_C'}{2\mu}\,\Big[\,V_{cc,h}+
V_{cn,h}\,g_{0n}^{}\,V_{nc,h}\,\Big]\,\varphi_{n\ell}^{},
\ee
which gives, using Eqs. (\rf{4e3}):
\bea \lb{8e10}
\mathcal{E}_{hvpol,n\ell}^{(2)}&=&
E_{hvpol,n\ell}^{(2)}-\frac{i}{2}\Gamma_{hvpol,n\ell}^{(2)}\nonumber \\
&=&2\Big(\varphi_{n0}^{\dagger}\,V_{cc,vpol}\,
\frac{G_C'}{2\mu}\,\Big)\,\frac{1}{\varphi_{n0}^{\dagger}(0)}
\Big(\,E_{h,n0}^{(1)}-\frac{i}{2}\Gamma_{h,n0}^{(1)}\,\Big)\,
\delta_{\ell 0}^{}.
\eea
Defining
\be \lb{8e11}
\delta_{hvpol,n0}^{(2)}\equiv \Big(\varphi_{n0}^{\dagger}\,V_{cc,vpol}
\,\frac{G_C'}{2\mu}\,\Big)\,\frac{1}{\varphi_{n0}^{\dagger}(0)},
\ee
one has:
\be \lb{8e12}
\frac{E_{hvpol,n0}^{(2)}}{E_{h,n0}^{(1)}}=2\delta_{hvpol,n0}^{(2)},
\ \ \ \ \ \frac{\Gamma_{hvpol,n0}^{(2)}}{\Gamma_{h,n0}^{(1)}}=
2\delta_{hvpol,n0}^{(2)}.
\ee
\par
The above correction is finite. Using the expression of the vacuum 
polarization potential $V_{cc,vpol}$ \cite{lp} and replacing $G_C'$ by 
a sum of contributions of intermediate states (discrete and continuous), 
one finds:
\be \lb{8e13}
\delta_{hvpol,10}^{(2)}=0.30\alpha\simeq 0.002,\ \ \ \ \
\delta_{hvpol,20}^{(2)}=0.28\alpha=0.002. 
\ee
\par
On comparing the orders of magnitude of the individual contributions
of $V_{cc,h}$ and $V_{cc,vpol}$ from results (\rf{8e8}) and (\rf{8e13}),
one deduces that the ratio of the contributions of the latter to the
former is of the order of 1/4; this implies that the contribution
of the quadratic piece in $V_{cc,vpol}$ is in the ratio of 1/16 with
respect to the quadratic piece of $V_{cc,h}$ and hence can be
neglected.
\par

\section{Summary of results} \lb{s9}
\setcounter{equation}{0}

We calculated the main corrections to the lowest-order formulas
of the energy shift and decay width of the $\pi K$ atom. At
lowest order, the energy shift and the decay width are given by
the formulas obtained by Deser \textit{et al.} (Sec. \rf{s4}).
They are expressed in terms of the $S$-wave scattering lengths 
of the strong interaction $\pi K\rightarrow \pi K$ scattering 
amplitudes, taken in the isospin symmetry limit; they are 
designated by $E_{h,n0}^{(1)}$ and $\Gamma_{h,n0}^{(1)}$, where
$n$ is the principal quantum number and $\ell$ (here equal to zero)
the orbital quantum number. 
\par
The main corrections, of order $\alpha$,  that arise are the following:
\par
1) Pure electromagnetic interaction effects beyond the Coulomb potential,
designated by $E_{elm,n\ell}^{(1)}$ (Sec. \rf{s5}) in the bound state
with quantum numbers $n$ and $\ell$. They contribute only to the real 
energy shift.
\par
2) Electromagnetic radiative corrections as well as isospin symmetry
breaking corrections to the strong interaction scattering amplitudes,
the relative amounts of which with respect to the lowest-order results
are designated by $\delta_{cc,h\gamma}^{(1)}$ and 
$\delta_{cn,h\gamma}^{(1)}$, according to the charged-charged
($\pi^-K^+\rightarrow \pi^-K^+$) and chargrd-neutral 
($\pi^-K^+\rightarrow \pi^0K^0$) channels (Sec. \rf{s6}).
\par
3) Corrections coming from second-order perturbation theory of the
expansion of the bound state energies (Sec. \rf{s8}). They involve 
strong interaction type correlations, represented by the relative amount 
$\delta_{hh,n0}^{(2)}$ with respect to the lowest-order results,
and strong interaction-vacuum polarization type correlations, represented 
by the relative amount $\delta_{hvpol,n0}^{(2)}$.
\par
The real energy shift, $\Delta E$, and the decay width, $\Gamma$,
including the $O(\alpha)$ corrections, take the following expressions:
\bea
\lb{9e1}
\Delta E_{n0}&=&-2\mu^2\frac{\alpha^3}{n^3}\,\big(a_0^++a_0^-\big)\,
\Big(\,1+\delta_{cc,h\gamma}^{(1)}+\delta_{hh,n0}^{(2)}+
2\delta_{hvpol,n0}^{(2)}\,\Big)+E_{elm,n0}^{(1)},\\
\lb{9e2}
\Delta E_{n1}&=&E_{elm,n1}^{(1)},\\
\lb{9e3}
\Gamma_{n0}&=&8p_{n0}^*\mu^2\frac{\alpha^3}{n^3}\,\big(a_0^-\big)_{}^2\, 
\Big(\,1+2\delta_{cn,h\gamma}^{(1)}+2\delta_{hh,n0}^{(2)}+ 
2\delta_{hvpol,n0}^{(2)}\,\Big),
\eea
where the $S$-wave scattering lengths $a_0^+$ and $a_0^-$ are those
of the pure strong interaction theory taken in the isospin symmetry
limit, $\mu$ the reduced mass of $\pi^-$ and $K^+$ and $p_{n0}^*$ 
the c.m. momentum of the neutral mesons after the decay of the bound
state with quantum numbers $(n,\ell=0)$.
\par
The numerical values of the various corrective terms for the first 
three bound states are summarized in Table \rf{9t1}.
\begin{table}[ht]
\bc
\begin{tabular}{|c|c|c|c|c|c|}
\hline
 $n$ $\ell$ & $\delta_{cc,h\gamma}^{(1)}$ &  
$\delta_{cn,h\gamma}^{(1)}$ & $\delta_{hh}^{(2)}$ &
$\delta_{hvpol}^{(2)}$ & $E_{elm}^{(1)}$ (eV)\\
\hline
 1 0 & $0.011\pm 0.032$ & $0.012\pm 0.007$ & 
0.009 & 0.002 & $-2.66$ \\
 2 0 & $0.011\pm 0.032$ & $0.012\pm 0.007$ & 
0.008 & 0.002 & $-0.32$ \\ 
 2 1 &  $\ 0.00$ & $\ 0.00$ & 0.000 & 0.000 & $-0.03$ \\
\hline
\end{tabular}
\ec
\caption{Numerical values of the correction terms to the lowest-order
formulas for the first three bound states.}
\lb{9t1}
\end{table}
\par
The uncertainties come mainly from the low energy constants of
ChPT in the presence of electr\-magne\-tism which are poorly known 
and are taken into account through their order of magnitude. A more
accurate evaluation of them, for instance by means of specific
models, would considerably reduce those uncertainties. 
\par
The decay width of the ground state and the energy
splitting between the $2P$ and $2S$ states take the following forms
in the presence of the $O(\alpha)$ corrections:
\bea
\lb{9e4}
& &\Gamma_{10}=8p_{10}^*\mu^2\alpha^3\,\big(a_0^-\big)_{}^2\,
\big(1+0.046\pm 0.014\big),\\
\lb{9e5}
& &\big(E_{21}-E_{20}\big)=
\frac{1}{4}\mu^2\alpha^3\,\big(a_0^++a_0^-\big)\,
\big(1+0.023\pm 0.032\big)+0.29\,(\mathrm{eV}).
\eea
These formulas allow one to extract from the experimental results
on the decay width and the energy splitting the values of the 
strong interaction scattering lengths $a_0^-$ and $a_0^+$. We observe
that the uncertainties in the corrective terms are much smaller in
the decay width than in the energy splitting. This means that the
decay width measurement will give us a more precise value for the 
scattering length $a_0^-$ than the measurement of the energy splitting
for the combination $(a_0^++a_0^-)$. On the other hand, the
latter quantity is sensitive, through its dependence on the low energy 
constants $L_4$ and $L_6$, to the Zweig rule violating effects \cite{gl2}
and therefore its precise knowledge is of crucial importance for the
understanding of the chiral symmetry breaking mechanism and of the 
dynamical role of the strange quark \cite{mous,dggs}. A precise 
knowledge of $a_0^-$, which mainly depends on the low energy constant
$L_5$ \cite{bkm}, allow us to have a better insight into the ratio 
$F_K/F_{\pi}$ of the kaon and pion weak decay constants \cite{gl2}.
\par 

\vspace{0.25 cm}
\noindent
\textbf{Acknowledgements}
\par
One of us (H.S.) thanks L. Nemenov for stimulating discussions on
hadronic atoms and informations on future projects about $\pi K$ atoms,
and B. Moussallam for discussions on $\pi K$ scattering amplitudes. 
Feynman diagrams have been drawn with the package Axodraw: 
J. A. M. Vermaseren, Comp.\ Phys.\ Comm.\ 83 (1994) 45.
This work was supported in part by the European Union network EURIDICE under 
EU RTN Contract CT2002-0311.
\par

\appendix
\renewcommand{\theequation}{\Alph{section}.\arabic{equation}}

\section{Three- and four-dimensional integrals} \lb{ap1}
\setcounter{equation}{0}

Integrals are calculated with dimensional regularization, with
dimension $d$ close to 4. $\overline \mu$ is the
mass scale of the $d$-dimensional theory.
\par
For the various integrals with two, three or four propagators, we use
a notation similar to that of Brown and Feynman \cite{bf}. In an
elastic two-particle scattering process, we designate by $p_1$ and 
$p_2$ the incoming particle momenta, with masses $m_1$ and $m_2$,
respectively, and by $p_1'$ and $p_2'$ the outgoing
particle momenta, with the total momentum $P=(p_1+p_2)=(p_1'+p_2')$ and 
the momentum transfer $q=(p_1-p_1')=(p_2'-p_2)$; the Mandelstam variables 
are: $s=P^2$, $t=q^2$, $u=(p_1-p_2')^2$.
\par
For the propagators, we use simplified notations,
where $k$ is the loop momentum variable:
\bea \lb{ae2}
\frac{1}{(1)}&=&\frac{i}{(p_1-k)^2-m_1^2+i\varepsilon},\ \ \ \ \ 
\frac{1}{(2)}=\frac{i}{(p_2+k)^2-m_2^2+i\varepsilon},\nonumber \\
\frac{1}{(-1')}&=&\frac{i}{(-p_1'-k)^2-m_1^2+i\varepsilon},\ \ \ \ \ 
\frac{1}{(-2')}=\frac{i}{(-p_2'+k)^2-m_2^2+i\varepsilon},\nonumber \\
\frac{1}{(0)}&=&\frac{-i}{k^2+i\varepsilon},\ \ \ \ \ \ \ \ 
\frac{1}{(3)}=\frac{-i}{(q-k)^2+i\varepsilon}.
\eea
The definitions of the integrals are:
\bea \lb{ae3}
J&=&\int\,[dk]\,\frac{1}{(1)(2)(0)(3)},\ \ \ \ \
F=\int\,[dk]\,\frac{1}{(1)(2)(3)},\ \ \ \ \ 
H=\int\,[dk]\,\frac{1}{(1)(2)(0)},\nonumber \\
G^{(1)}&=&\int\,[dk]\,\frac{1}{(1)(0)(3)},\ \ \ \ \ 
G^{(2)}=\int\,[dk]\,\frac{1}{(2)(0)(3)},\ \ \ \ \ 
A=\int\,[dk]\,\frac{1}{(1)(2)},
\eea
where $[dk]=\overline\mu^{4-d}d^dk/(2\pi)^d$.
Vector and tensor generalizations of these integrals correspond to
the cases where momenta $k_{\mu}$ or $k_{\mu}k_{\nu}$ appear in the
numerator of the integrands.
\par
Crossed diagrams involve integrals where $(2)$ is replaced by $(-2')$.
The corresponding integrals are defined as:
\be \lb{ae4}
J(1,-2')=\int\,[dk]\,\frac{1}{(1)(-2')(0)(3)},
\ee
etc.
\par
Constraint diagrams involve three-dimensional integrals which result
from the $s$-chan\-nel four-dimensional integrals by the replacement of
the two propagators of the incoming particles by the single effective
propagator $g_0^{}$ [Eq. (\rf{2e5})]. The corresponding integrals are,
in the c.m. frame:
\bea \lb{ae5} 
J^C&=&-\frac{2\pi}{2\sqrt{s}}\,\int\,[dk]\,\delta(k_0)\,
\frac{1}{(1)(0)(3)},\ \ \ \ \
F^C=-\frac{2\pi}{2\sqrt{s}}\,\int\,[dk]\,\delta(k_0)\,
\frac{1}{(1)(3)},\nonumber \\
H^C&=&-\frac{2\pi}{2\sqrt{s}}\,\int\,[dk]\,\delta(k_0)\,
\frac{1}{(1)(0)},\ \ \ \ \ A^C=-\frac{2\pi}{2\sqrt{s}}\,\int\,[dk]\,
\delta(k_0)\,\frac{1}{(1)}.
\eea
The factor $1/(2\sqrt{s})$ that has been incorporated in the definitions 
takes account of the fact that constraint diagrams contain one more
scattering amplitude than the four-dimensional integrals and the latter
is defined with the factor $i/(2\sqrt{s})$ [Eq. (\rf{2e2})]; the factor
$i$ is now contained in the definition of the meson propagator; the minus 
sign is reminiscent of a similar sign in front of $g_0^{}$ in Eq. (\rf{2e3}).
\par
The three-dimensional integral of the effective propagator $g_0^{}$
[Eqs. (\rf{2e4})-(\rf{2e5})] is, in the vicinity of $d=4$:
\bea \lb{ae1}
i2\sqrt{s}\,A^C&=&\overline\mu^{4-d}\int\,\frac{d^{d-1}k}{(2\pi)^{d-1}}\,
\frac{1}{b_0^2(s)-\mathbf{k}^2+i\varepsilon}\nonumber \\
&=&\frac{1}{4\pi}\sqrt{-b_0^2(s)}\,
\Big[\,1+(\frac{d}{2}-2)\Big(-\psi(1)-2+
\ln\big(\frac{-4b_0^2(s)}{4\pi\overline\mu^2}\big)\Big)\,\Big],
\eea
where $-b_0^2(s)$ is taken positive and $\psi$ is the logarithmic
derivative of the Gamma function. 
Analytic continuation to positive values of $b_0^2(s)$ is done with
the replacement $\sqrt{-b_0^2(s)}\rightarrow -i\sqrt{b_0^2}(s)$.
\par
The integral $A$ is:
\be \lb{ae17}
A=\frac{i}{16\pi^2}\,\Big[\,\frac{2}{d-4}-\psi(1)-2
+\ln\big(\frac{m_1m_2}{4\pi\overline\mu^2}\big)
+\frac{(m_1^2-m_2^2)}{2s}\ln\big(\frac{m_1^2}{m_2^2}\big)+Q(s)\,\Big],
\ee
with
\be \lb{ae18}
Q(s)=\left\{
\ba{l}
+\frac{\sqrt{4sb_0^2(s)}}{s}\Big[\ln\Big(\frac{\sqrt{s-(m_1-m_2)^2}
+\sqrt{s-(m_1+m_2)^2}}{\sqrt{s-(m_1-m_2)^2}-\sqrt{s-(m_1+m_2)^2}}\Big)
-i\pi\Big],\ \ \ \ (m_1+m_2)^2<s,\\
+\frac{\sqrt{-4sb_0^2(s)}}{s}\Big[\pi-\arctan\Big(\frac{\sqrt{-4sb_0^2(s)}}
{s-(m_1^2+m_2^2)}\Big)\Big],\ \ \ \ (m_1-m_2)^2<s<(m_1+m_2)^2,\\
-\frac{\sqrt{4sb_0^2(s)}}{s}\ln\Big(\frac{\sqrt{(m_1+m_2)^2-s}+\sqrt{
(m_1-m_2)^2-s}}{\sqrt{(m_1+m_2)^2-s}-\sqrt{(m_1-m_2)^2-s}}\Big),\ \ \ \
s<(m_1-m_2)^2.
\ea
\right.
\ee
The integral of $\widetilde G_0$, entering in Eqs. (\rf{2e5a}),
(\rf{6e1}) and (\rf{7e1}), is equal to $-i2\sqrt{s}A$.
\par
The integral $J$ is equal, on the mass shell, to:
\bea \lb{ae6}
J&=&-\frac{i}{16\pi^2}\frac{1}{t\sqrt{-b_0^2(s)s}}\,\Big[\,\frac{2}{d-4}
-\psi(1)+\ln\big(\frac{-t}{4\pi\overline\mu^2}\big)\,\Big]\,
\nonumber \\
& &\ \ \ 
\times \Big[\,\arctan\Big(\frac{s+m_1^2-m_2^2}{2\sqrt{-b_0^2(s)s}}\Big)+
\arctan\Big(\frac{s-m_1^2+m_2^2}{2\sqrt{-b_0^2(s)s}}\Big)\,\Big].
\eea
Its threshold expansion is obtained by taking into account the facts that
$|t|\ll s\simeq (m_1+m_2)^2$, $|b_0^2(s)|\ll s$:
\be \lb{ae7}
J=-\frac{i}{16\pi^2}\frac{1}{t}\,\Big[\,\frac{2}{d-4}
-\psi(1)+\ln\big(\frac{-t}{4\pi\overline\mu^2}\big)\,\Big]\,
\Big[\,\frac{\pi}{\sqrt{-b_0^2(s)s}}-\frac{1}{m_1m_2}\,\Big]
+O(\alpha^3\ln\alpha^{-1}),
\ee
where the order of magnitudes are evaluated with the counting rules
of the QED bound states. The latter expansion can also be obtained
with the threshold expansion method of Ref. \cite{bs}. The singularity in
$1/\sqrt{-b_0^2(s)s}$ is produced by the potential momenta, while the term
in $1/(m_1m_2)$ is produced by the soft momenta. Ultrasoft momenta do
not contribute at leading orders, while the hard momenta contribute
at $O(\alpha^3)$.
\par
The expression of $J(1,-2')$ is obtained from that of $J$ by the
replacement of the variable $s$ by $u$. Its threshold expansion is
obtained by noticing that 
$u\simeq (m_1-m_2)^2-t-b_0^2(s)(m_1+m_2)^2/(m_1m_2)$:
\be \lb{ae8}
J(1,-2')=-\frac{i}{16\pi^2}\frac{1}{t}\,\Big[\,\frac{2}{d-4}
-\psi(1)+\ln\big(\frac{-t}{4\pi\overline\mu^2}\big)\,\Big]\,
\frac{1}{m_1m_2}+O(\alpha^3\ln\alpha^{-1}).
\ee 
The potential momenta do not contribute at leading order to $J(1,-2')$
and the leading term comes now from the soft momenta.
\par
The three-dimensional integral $J^C$ is equal to:
\be \lb{ae9}
J^C=\frac{i}{16\pi^2}\frac{1}{t}\,\Big[\,\frac{2}{d-4}
-\psi(1)+\ln\big(\frac{-t}{4\pi\overline\mu^2}\big)\,\Big]\,
\frac{\pi}{\sqrt{-b_0^2(s)s}}.
\ee
\par
Taking now the sum of the integrals $J$, $J(1,-2')$ and $J^C$, we find
that $J^C$ cancels the potential momenta contribution of $J$, while the
soft momenta contributions of $J$ and $J(1,-2')$ cancel each other:
\be \lb{ae10}
J+J(1,-2')+J^C=O(\alpha^3\ln\alpha^{-1}).
\ee
\par
A similar type of analysis also applies to the sum of the three diagrams
(b), (c) and (d) of Fig. \rf{5f1} arising in QED. Here, the couplings
being vector-like, one first decomposes the various integrals into a
tensor sum, involving integrals of the types $J$, $F$, $H$, $G$, etc. and
their vector and tensor associates. Taking into account the $\alpha^2$
factor coming from the couplings, the sum of all these contributions
vanishes up to order $\alpha^4$. (Notice that the constraint diagrams
should be calculated as three-dimensional integrals involving $g_0^{}$ 
and one or two on-mass shell one-photon exchange diagrams.)
\par 
The above result does not remain true in an off-mass shell formalism.
Here, the scattering amplitude is no longer gauge invariant and slight
differences arise. The sum (\rf{ae10}) yields now a $O(\alpha)$ term
and generally the sum of the three previous diagrams behaves as
$O(\alpha^3)$ \cite{js2}. The latter term, which is spurious, is cancelled
by a higher-order diagram. Nevertheless, in the Fried-Yennie gauge 
\cite{fy}, the same results as in the on-mass shell formalism occur.
\par
Integrals $F$ and $H$ appear also in electromagnetic radiative corrections
to the strong interaction. Integral $H$ appears in diagram (c) of Fig.
\rf{6f1}; diagram (b) corresponds to its crossed diagram involving
$H(1,-2')$; diagram (d) involves $H^C$; diagram (a) involves $H$ in the 
$t$-channel. $H$ has the following threshold expansion:
\bea \lb{ae11}
H&=&-\frac{1}{32\pi^2}\,\frac{\pi}{\sqrt{-b_0^2(s)s}}\,\Big[\,
\frac{2}{d-4}-\psi(1)+\ln\big(\frac{-4b_0^2(s)}{4\pi\overline\mu^2}\big)
\,\Big]\nonumber \\
& &-\frac{1}{32\pi^2 s}\,\bigg\{\,\Big[\,\frac{2}{d-4}-\psi(1)+
\ln\big(\frac{s}{4\pi\overline\mu^2}\big)+2\,\Big]\,
\frac{1}{\beta_1\beta_2}-2\,\Big[\,\frac{1}{\beta_1}\ln(\frac{1}{\beta_1})
+\frac{1}{\beta_2}\ln(\frac{1}{\beta_2})\,\Big]\,\bigg\}\nonumber \\
& &+O(\alpha^4\ln\alpha^{-1}),\ \ \ \ \ 
\beta_1=\frac{1}{2}\big(1+\frac{m_1^2-m_2^2}{s}\big),
\ \ \ \ \ \beta_2=\frac{1}{2}\big(1-\frac{m_1^2-m_2^2}{s}\big),
\eea
where the dominant singularity comes from the potential momenta and
the next-to-leading terms from the hard momenta. (The soft and
ultrasoft momenta do not contribute at leading orders.)
\par
The crossed integral to $H$ is:
\bea \lb{ae12}
H(1,-2')&=&\frac{1}{32\pi^2 s}\,\bigg\{\,\Big[\,\frac{2}{d-4}-\psi(1)+
\ln\big(\frac{s}{4\pi\overline\mu^2}\big)+2\,\Big]\,
\frac{1}{\beta_1\beta_2}\nonumber \\
& &+\frac{2}{(\beta_1-\beta_2)}\,
\Big[\,\frac{1}{\beta_1}\ln(\frac{1}{\beta_1})
-\frac{1}{\beta_2}\ln(\frac{1}{\beta_2})\,\Big]\,\bigg\}
+O(\alpha^4\ln\alpha^{-1}),
\eea
where only hard momenta contribute.
The integral entering in the $t$-channel vertex function (diagram (a)
of Fig. \rf{6f1}) is $H(1,-1';m_1,m_1)$:
\be \lb{ae13}
H(1,-1';,m_1,m_1)=\frac{1}{32\pi^2 m_1^2}\,\Big[\,\frac{2}{d-4}-\psi(1)+
\ln\big(\frac{m_1^2}{4\pi\overline\mu^2}\big)\,\Big]
+O(\alpha^4\ln\alpha^{-1}).
\ee
It can also be obtained from the result (\rf{ae12}), by taking in it
the limit $m_2\rightarrow m_1$.
\par
$H^C$ is equal to the opposite of the contribution of the potential
momenta in $H$:
\be \lb{ae14}
H^C=\frac{1}{32\pi^2}\,\frac{\pi}{\sqrt{-b_0^2(s)s}}\,\Big[\,
\frac{2}{d-4}-\psi(1)+\ln\big(\frac{-4b_0^2(s)}{4\pi\overline\mu^2}\big)
\,\Big].
\ee
\par
If $\Sigma(p^2,m^2)$ is the meson electromagnetic self-energy, taken for
the moment with scalar couplings, then, after a mass-shell 
renormalization, it is the quantity 
$\frac{1}{2}\frac{\partial \Sigma}{\partial m^2}|_{p^2=m^2}$ that
multiplies in lowest order the strong interaction vertex. This yields
$-1/2$ of the value of the $t$-channel form factor at $t=0$ [Eq.
(\rf{ae13})]. The self-energy contributions of the two external 
mesons of that form factor then cancel the latter completely at
leading order. Similarly, the sum of $H$, $H(1,-2')$ and $H^C$ yields
a finite $O(\alpha^3)$ term.
\par
The above results should be completed by incorporating the vector
coupling of the photon. The latter does not change the leading behavior 
of $H^C$. Concerning the four-dimensional integrals, the only 
modifications are through the hard momenta contributing with finite
$O(\alpha^3)$ effects. (Loop momenta in the numerators improve the
infrared behavior of the object.) Taking also into account the factor
$\alpha$ coming from the photon couplings, the sum of all 
contributions reduces to a finite $O(\alpha^4)$ term.
\par
Diagrams where the photon is emitted from the vertex are not infrared
singular and give contributions of order $\alpha^4$.
\par
In the case of the charged-neutral channel (process 
$\pi^-K^+\rightarrow \pi^0K^0$), the integrals of the $t$-channel and 
$u$-channel form factors are absent. In that case, the cancellations
occur between the sum $H+H^C$ and the contributions of the self-energies
of the two external charged mesons, taking into account the vector
coupling of the photon. The result is again a finite $O(\alpha^4)$ term.
\par
The integral that contributes to the dominant part of the constraint 
diagram (d) of Fig. \rf{6f2} is denoted $I^{CC}$; it is:
\bea \lb{ae15}
I^{CC}&=&\frac{1}{2\sqrt{s}}\,\overline\mu^{2(4-d)}\,\int\,   
\frac{d^{d-1}k}{(2\pi)^{d-1}}\frac{d^{d-1}k'}{(2\pi)^{d-1}}\,
\frac{i}{b_0^2(s)-\mathbf{k}^2+i\varepsilon}\,     
\frac{i}{b_0^2(s)-\mathbf{k}^{\prime 2}+i\varepsilon}\,
\frac{(-i)}{-(\mathbf{k}-\mathbf{k}')^2+i\varepsilon}\nonumber \\
&=&\frac{i}{64\pi^2\sqrt{s}}\,\Big(\frac{1}{d-4}-\psi(1)
-1+\ln\big(\frac{-4b_0^2(s)}{4\pi\overline\mu^2}\big)\Big).
\eea
\par
The integral that contributes to the dominant part of diagram (b) of
Fig. \rf{6f2}, denoted $I^C$ is calculated in the following way. One
first calculates the four-dimensional integral on the mass-shell; the
latter is then integrated three-dimensionnally, by extending
eventually the domain of validity of the momentum transfer squared $t$.
The four-dimensional integral on the mass-shell is nothing but $H$
[Eq. (\rf{ae11})]. The latter is independent of $t$. Hence the 
three-dimensional integration reduces to that of $ig_0^{}$, given by
Eq. (\rf{ae1}) times $i$. The result is therefore the product of the 
right-hand sides of Eqs. (\rf{ae11}) and (\rf{ae1}), times $i$; since 
the integral of $g_0^{}$ is of order $\sqrt{-b_0^2(s)}$, it is sufficient 
to retain the dominant singularity of $H$. One finds the opposite value of 
$I^{CC}$ [Eq. (\rf{ae15})]. Hence the sum of the three constraint diagrams 
of Fig. \rf{6f2} is given by the opposite value of $I^{CC}$; the finite
logarithmic part of it cancels a similar term in the diagram (a) of Fig.
\rf{6f2} \cite{bs}. As we emphasized in Secs. \rf{s7} and \rf{s8}, the 
infinite part of $I^{CC}$ is cancelled by a similar term present in 
second-order of the perturbation theory expansion of the bound state 
energy. The finite part that has been retained in the right-hand side 
of Eq. (\rf{8e7}), second term, corresponds to the contribution of the 
factor $-b_0^2(s)$ of the logarithm.
\par   
The integrals $G$ [Eq. (\rf{ae3})] are infrared finite and receive
contributions at leading order from potential and soft momenta with
the ratio $(2)/(-1)$:
\be \lb{ae16}
G^{(a)}=-\frac{1}{32m_a}\frac{1}{\sqrt{-t}}+O(\alpha^3\ln\alpha^{-1}),
\ \ \ \ \ a=1,2.
\ee
\par

\par

\end{document}